\renewcommand\k{\kappa}

\newcommand{\diracslash}[1]{#1\llap{/\kern2pt}}

\newcommand{\be}{\begin{equation}}
\newcommand{\ee}{\end{equation}}
\newcommand{\bea}{\begin{eqnarray}}
\newcommand{\eea}{\end{eqnarray}}
\newcommand{\ba}[1]{\begin{array}{#1}}
\newcommand{\ea}{\end{array}}

\newcommand{\bt}{\begin{tabular}}
\newcommand{\et}{\end{tabular}}

\newcommand{\beas}{\begin{eqnarray*}}
\newcommand{\eeas}{\end{eqnarray*}}

\documentclass[preprint,prd,aps,floats,nofootinbib,showpacs,floatfix]{revtex4}
\usepackage{graphicx}
\begin{document}

\title{ $\rho$-meson spectral function in hot nuclear matter}
\author{P.C.Raje Bhageerathi}
\email{pc_raje@yahoo.com}
\affiliation{Department of Physics, Indian Institute of Technology, Delhi,
Hauz Khas, New Delhi -- 110 016, India}

\author{Amruta Mishra}
\email{amruta@physics.iitd.ac.in,mishra@th.physik.uni-frankfurt.de}
\affiliation{Department of Physics, Indian Institute of Technology, Delhi,
Hauz Khas, New Delhi -- 110 016, India}

\begin{abstract}
We study the $\rho$-meson spectral function in hot 
nuclear matter by taking into account the isospin-symmetric pion
and the nucleon loops within the quantum hadrodynamics (QHD) model as well as using an effective 
chiral SU(3) model. The spectral function of the $ \rho $ meson is studied in the mean field approximation (MFA) as well as in the  relativistic Hartree (RHA) approximation. The inclusion of the nucleon loop
considerably changes the $\rho$-meson spectral function. Due to a larger mass drop of $ \rho $ meson in the RHA, it is seen that the spectral function shifts
towards the low invariant mass region, whereas in the MFA the spectral function is seen to be slightly shifted towards 
the high mass region. Moreover, while the spectral function is observed to be sharper with the nucleon-antinucleon polarization in RHA, the spectral function is seen to be broader in the MFA.

\end{abstract}
\maketitle

\def\bfm#1{\mbox{\boldmath $#1$}}

\section{Introduction}
The study of strongly interacting nuclear matter
under extreme conditions has attracted a lot of 
attention during the recent years, both theoretically and experimentally. The properties of hadrons
at high temperatures and densities are quite 
different from the properties of hadrons in vacuum. The quantum hadrodynamics
QHD-I model (Walecka model) and its extensions have been 
widely used to discuss the properties of the  
nuclear matter and finite nuclei \cite{Rapp, Brown, Wal1, Ser1, Ser2}. 
The ongoing relativistic heavy ion collision experiments,
at the high energy accelerators SPS, CERN, Switzerland;
SIS, GSI, Germany; RHIC, BNL, USA; LHC, CERN, Switzerland, 
and the compressed baryonic matter (CBM) experiments planned at the
future facilities at GSI, Germany, 
are intended to probe matter at high temperatures and densities.
The hadrons modified in the hot and dense hadronic medium
resulting from heavy ion collision experiments, affect the
experimental observables. E.g.,the dilepton spectra
observed from heavy ion collision experiments at the SPS
\cite{ceres,helios} are attributed to the
medium modifications of vector mesons 
\cite{Brat1,CB99,vecmass,dilepton,liko,zschocke,haya,weise10},
and can not be explained by vacuum hadronic properties.

It is predicted that in ultra-relativistic heavy-ion collisions 
quark-gluon plasma may be produced and the spontaneously broken
chiral symmetry may be restored. Amongst the proposed signals for 
detecting the quark-hadron phase transition, dileptons and photons
are considered to be the cleanest ones because they do not interact 
with the hadronic
medium and emerge from the heavy ion collision experiments almost 
undisturbed \cite{Shu, KK, IT}. Also considering the 
fact that the light vector mesons can directly decay to dilepton pairs,
the study of the $\rho$ meson in the medium is interesting because of its
relatively large decay width as compared to those of $\omega$ and 
$\phi$ mesons. 
\cite{Teo1, Teo2, Song1, Gale1, Hat1, Leu, Kling}.  

One of the hot topics in recent years is discussing the low invariant 
mass dilepton production in heavy-ion collisions. The dilepton 
distribution is related with the spectral function of $\rho$ meson
\cite{Gale1, Wel}. In the present work, the $\rho$ meson spectral 
function is calculated accounting for effects of the pion as well as the nucleon
loops. The effect of modification of the nucleon mass on the 
$\rho$ meson spectral function is discussed within the framework of
both the Walecka model and a chiral SU(3)  model.
In both the models, the effective nucleon mass is modified with density
and temperature and this in turn modifies the
$\rho$ meson spectral function.

We organize the paper as follows: Section II gives a brief
description of the QHD-I (Walecka) model and the chiral SU(3) model used in the 
present investigation. In Section III, we discuss the nucleon
properties in hot and dense matter in both of these models. Section IV discusses
the effects of finite temperature and density on the spectral  
function of the $\rho$ meson. Section V contains the results
and discussion and in section VI, we summarize our main findings of the present investigation and discuss possible outlook.

\section{ The hadronic models }
\subsection{QHD-I (Walecka) model}

The Lagrangian density for the model is given by \cite{Ser1, Ser2}
\begin{eqnarray}
L = & & \bar{\psi}\left[ \gamma_{\mu}\left( i\partial^{\mu}- g_{\omega}\omega^{\mu}\right) -\left( M-g_{\sigma}\sigma\right) \right] \psi  + \frac{1}{2}\left( \partial_{\mu}\sigma\partial^{\mu}\sigma-m_{\sigma}^{2}\sigma^{2}\right) -\frac{1}{3!}\kappa\sigma^{3} - \frac{1}{4!}\lambda\sigma^{4}\nonumber\\
& & -\frac{1}{4}F_{\mu\nu}F^{\mu\nu} +\frac{1}{2}m_{\omega}^{2}\omega_{\mu}\omega^{\mu} + \delta L \nonumber\\
\end{eqnarray}
where $\psi$ is the nucleon field, $\sigma$ is the neutral scalar meson field
and $\omega$ is the isoscalar vector field.
$F^{\mu\nu}= \partial_\mu \omega^\nu - \partial_\nu \omega^\mu  $ is the field tensor for the vector meson, $ \omega $ 
and $ \delta L $ contains counterterms used for renormalization. 
The parameters $ M, g_\sigma, g_\omega, m_\sigma, m_\omega, \kappa $,
and $\lambda  $ are phenomenological constants that are determined
from the nuclear matter saturation properties. 

In the QHD-I (Walecka model), the nucleons interact through the 
exchange of $\sigma$ and $\omega$ mesons.  The $\sigma$ exchange 
gives the attractive force while the $\omega$ exchange attributes to the
repulsive interaction between the nucleons. We use the QHD-I model to obtain the effective
nucleon mass $ M_N^* $ and effective chemical potential $ \mu^* $  in the hot and dense nuclear matter and use the 
obtained $ M_N^* $ and $ \mu^* $ 
in mean field or relativistic Hartree approximation for investigating the in-medium properties of the vector mesons. 

\subsection{The hadronic chiral SU(3)$ \times $ SU(3) model}

The effective hadronic chiral Lagrangian density used in the present work is given as
\begin{equation}
L=L_{kin}+\sum_{W=X,Y,V,A,u} L_{BW} +L_{vec}+L_{0}+L_{SB}
\end{equation}
Equation (2) corresponds to a relativistic model of baryons and mesons adopting a nonlinear realization of chiral symmetry \cite{Wein, Cole, Bar} and broken scale invariance as a description of the hadronic matter. Here, $L_{kin}$ is kinetic energy term, $L_{BW}$ is the baryon-meson interaction term in which the baryons-spin-0 meson interaction term generates the baryon masses. $L_{vec}$  describes the dynamical mass generation of the vector mesons via couplings to the scalar mesons and contains additionally quartic self-interactions of the vector fields. $L_{0}$ contains the meson-meson interaction terms as well as a scale invariance breaking logarthimic potential. $L_{SB}$ describes the explicit chiral symmetry breaking. 
The baryon-scalar meson interactions generate the baryon masses
and the parameters corresponding to these interactions are adjusted 
so as to obtain the baryon masses as their experimentally measured 
vacuum values. For the baryon-vector meson interaction terms, there
exist the $F$-type (antisymmetric) and $D$-type (symmetric) couplings.
Here we use the antisymmetric coupling \cite{paper3,isoamss1,isoamss2}
because, following the universality principle  \cite{saku69}
and the vector meson dominance model, one can conclude that
the symmetric coupling should be small.
Additionally we choose the parameters \cite{paper3,isoamss} so as to
decouple the strange vector field $
\phi_\mu\sim\bar{s} \gamma_\mu s $ from the nucleon,
corresponding to an ideal mixing between $\omega$ and $\phi$.
A small deviation of the mixing angle from the ideal mixing
\cite {dumbrajs,rijken,hohler1} has not been taken into
account in the present investigation. The Lagrangian densities corresponding to the interaction
for the vector meson, $ L_{vec}$, the meson-meson interaction
$ L_0$ and that corresponding to the explicit chiral symmetry breaking
$ L_{SB}$ have been described in detail in references
\cite{paper3,isoamss}.

To investigate the hadronic properties in the medium, we write
the Lagrangian density within the chiral SU(3) model 
and determine the expectation values
of the meson fields by solving the equations of motion of the scalar fields at finite temperature and density obtained by minimizing the thermodynamic potential.

\section{Nucleon properties in hot nuclear matter}

First we proceed to study the hadronic properties in the QHD-I
model. We study the nucleon properties in both the mean field and relativistic Hartree approximations (RHA).  
In the mean field approximation (MFA), the meson field 
operators can be approximated by their expectation values, which
are the classical fields. The expectation value of the scalar field,  $\sigma_0$ shifts the nucleon mass from $ M_{N} $ to
$ M_{N}^{*}=M_{N}- g_{\sigma} \sigma_{0} $ and the vector
field gives rise to an effective chemical potential,
$ \mu^{\ast} $ = $ \mu - \frac{g_{\omega N}^{2}\rho_{B}}{m_{\omega}}^{2} $.
The nucleon effective mass can be determined self-consistently by \cite{Ser1}
\begin{equation}
M_{N}-M_{N}^{\ast}  = -\left( \frac{g_{\sigma}^{2}}{m_{\sigma}^{2}}\right) \rho_{B}^{S}
\end{equation}
where the scalar density $ \rho_{B}^{S} $ of the nuclear matter is 
\begin{eqnarray}
\rho_{B}^{S} = \gamma_{N}\int\frac{d^{3}p}{(2\pi)^{3}} \frac{M_{N}^{*}}{E_{N}^{*}} (n_{N}(p) + \bar{n}_{N}(p)) 
\end{eqnarray}
with the spin-isospin degeneracy factor, $\gamma$ = 4 for symmetric nuclear matter.
The effective chemical potential $\mu^*$ is determined by
\begin{equation}
\mu^{\ast}=\mu-\frac{g_{\omega N}^{2}\rho_{B}}{m_{v}^{2}}
\end{equation}
with $ \rho_{B} $ as the baryon density, given by,
\begin{equation}
\rho_{B} = \gamma_{N}\int\frac{d^{3}p}{(2\pi)^{3}} (n_{N}(p) - \bar{n}_{N}(p))  
\end{equation} 
Here, $ n_{N} $ and $ \bar{n_{N}} $ are the thermal distribution functions given as
\begin{equation}
n_{N}(p) = \frac{1}{e^{(E_{N}^{*}-\mu^{*})/T}+1}  ,\nonumber\\
\bar{n}_{N}(p) = \frac{1}{e^{(E_{N}^{*}+\mu^{*})/T}+1}  .  
\end{equation}
where, $ E_{N}^{*} = \sqrt{p^{2}+(M_{N}^{*})^{^{2}}} $ is the single particle energy of the nucleon and T is the temperature.
One can solve the above coupled equations numerically to obtain
the effective nucleon mass $ M_{N}^{*} $ and the effective
chemical potential $ \mu^{*} $ for given density, $ \rho_{B} $ and temperature, T.  The RHA takes into account the vacuum 
fluctuation corrections to the mean field results.

Next we proceed to study the hadronic properties in the chiral
SU(3) model used in the present investigation \cite{Zsch1, Zsch2}. The Lagrangian density in the mean field approximation
is given as 
\begin{eqnarray}
L_{BX}+ L_{BV}  = & & -\bar{\psi_{N}}\left[ g_{N\omega}\gamma_{0}\omega +  M_{N}^{\ast}\right] \psi_{N} , \nonumber\\
L_{vec} = & & \frac{1}{2}m_{\omega}^{2}\frac{\chi^{2}}{\chi_{0}^{2}}\omega^{2} +  g_{4}^{4}\omega^{4} ,\nonumber\\
-L_{0} = & & \frac{1}{2}k_{0}\chi^{2}\left( \sigma^{2}+\zeta^{2}\right)- k_{1} \left( \sigma^{2}+\zeta^{2}\right)^{2} - k_{2}
\left( \frac{\sigma^{4}}{2}+\zeta^{4}\right) - k_{3}\chi\sigma^{2}\zeta \nonumber\\
& & +  k_{4}\chi^{4} +  \frac{1}{4}\chi^{4}\ln \frac{\chi^{4}}{\chi_{0}^{4}} - \frac{\delta}{3}\chi^{4}\ln \frac{\sigma^{2}\zeta}{\sigma_{0}^{2}\zeta_{0}} , \nonumber\\
-L_{SB} = & & \left( \frac{\chi}{\chi_{0}}\right) ^{2}\left[ m_{\pi}^{2}f_{\pi}\sigma + \left( \sqrt{2}m_{K}^{^{2}} f_{K} - \frac{1}{\sqrt{2}}m_{\pi}^{2}f_{\pi}\right) \zeta\right]  ,
\end{eqnarray}
where $ M_{N}^{\ast} $ = $ -g_{\sigma N} \sigma -g_{\zeta N}\zeta $ is the effective 
mass of the nucleon. The thermodynamical potential of the grand canonical
ensemble  $ \Omega $  per unit volume V at a given chemical potential 
$ \mu $ and temperature T can be written as
\begin{eqnarray}
\frac{\Omega}{V} = & & - L_{vec} - L_{0} - L_{SB} - \nu_{vac} + \gamma_{N}\int\frac{d^{3}p}{(2\pi)^{3}}E_{N}^{*}(p) (n_{N}(p) + \bar{n}_{N}(p)) \nonumber\\
& & - \gamma_{N}\int\frac{d^{3}p}{(2\pi)^{3}}\mu_{N}^{*} (n_{N}(p) - \bar{n}_{N}(p)) 
\end{eqnarray}
Here $ \gamma_{N} $ are the spin-isospin degeneracy factor, and $ \gamma_{N} $ = 4 
for symmetric nuclear matter. The $ n_{N}  $ and $ \bar{n}_{N} $ are the thermal 
distribution functions for the nucleon and the antinucleon given in terms
of the effective single particle energy, $ E_{N}^{*} $, and the effective chemical 
potential, $ \mu^{*} $, as given by equation (7). The mesonic field equations are 
determined by minimizing the thermodynamic potential.
We shall use the frozen glueball approximation $ \left( \chi = \chi_{0}\right) , $ since the dilaton field which simulates the gluon condensate changes very little in the medium. We then have 
coupled equations only for the fields $ \sigma,\zeta $  and  $ \omega $ 
as given by
\begin{eqnarray}
\frac{\partial\left( \Omega/V\right) }{\partial\sigma} = & & k_{0}\chi^{2}\sigma - 4k_{1}\left( \sigma^{2}+\zeta^{2}\right) \sigma - 2k_{2}\sigma^{3} - 2k_{3}\chi\sigma\zeta - 2\frac{\delta\chi^{4}}{3\sigma} \nonumber\\
& & + m_{\pi}^{2}f_{\pi} + \frac{\partial M_{N}^{\ast}}{\partial\sigma}\rho_{B}^{S} = 0,
\end{eqnarray}
\begin{eqnarray}
\frac{\partial\left(\Omega/V \right) }{\partial\zeta} = & & k_{0}\chi^{2}\zeta - 4k_{1}\left( \sigma^{2}+\zeta^{2}\right) \zeta - 4k_{2}\zeta^{3} - k_{3}\chi\sigma^{2} - \frac{\delta\chi^{4}}{3\zeta} \nonumber\\
& & + \frac{\partial M_{N}^{\ast}}{\partial\zeta}\rho_{B}^{S} + \left[ \sqrt{2}m_{K}^{2}f_{K}- \frac{1}{\sqrt{2}}m_{\pi}^{2}f_{\pi}\right]  = 0, 
\end{eqnarray}
\begin{eqnarray}
\frac{\partial\left( \Omega/V \right) }{\partial\omega} = & & - m_{\omega}^{2}\omega - 4g_{4}^{4}\omega^{3} + g_{N\omega}\rho_{B} = 0,
\end{eqnarray}
which have to be solved self-consistently to obtain the values of $ \sigma $, $ \zeta $ and $ \omega $ .      
Here $ \rho_{B}^{S} $  and  $ \rho_{B } $ are the scalar and vector densities
for the nuclear matter at finite temperature, T given by equations (4) and (6).

\section{Rho meson spectral function}

The most consistent approach for studying the hadronic matter produced
in ultra-relativistic heavy-ion collisions at high temperature and 
density is the finite temperature field theory \cite{Das, Bel}.
In the present calculation thermal effects enter through thermal nucleon
and pion loops. In Minkowski space, the self-energy of the $ \rho $
vector meson can be expressed as \cite{Gale1,Kap}
\begin{eqnarray}
\Pi^{\mu\nu}\left( k\right) & = & \Pi_{L}\left( k\right)P_{L}^{\mu\nu} + \Pi_{T}\left( k\right)P_{T}^{\mu\nu},
\end{eqnarray}
where $ k^{2} $ = $ k_{0}^{2}-|\vec{k}^{2}| $. The $ P_{L}^{\mu\nu} $ and $ P_{T}^{\mu\nu}  $
are the longitudinal and transverse projection tensors defined as
\begin{eqnarray}
P_{T}^{00}=P_{T}^{0i}=P_{T}^{i0}=0 ,\: P_{T}^{ij}=\delta^{ij}-\frac{k_{i}k_{j}}{|\vec{k}^{2}|} ,\: 
& P_{L}^{\mu\nu}=\frac{k^{\mu}k^{\nu}}{k^{2}}-g^{\mu\nu}-P_{T}^{\mu\nu}.
\end{eqnarray}
$ \Pi_{L} $ and $ \Pi_{T} $ are related to the components of 
the self-energy by
\begin{eqnarray}
\Pi_{L}\left( k\right) = -\frac{k^{2}}{|\vec{k}^{2}|}\Pi^{00}\left( k\right) , \nonumber\\ 
\Pi_{T}\left( k\right) = \frac{1}{2}\left( \Pi_{\mu}^{\mu}-\Pi_{L}\left( k\right) \right) .
\end{eqnarray}

The imaginary part of the retarded propagator is referred to as the 
spectral function and is related to the dilepton production. The study
of the $ \rho $ meson spectral function is attributed to calculating the
in-medium self-energy of the $ \rho $ meson.

\subsection{$ \rho NN $ interaction}

The contribution of nucleon excitations through nucleon-loop
to $ \rho $ self-energy is analyzed in terms of the effective
Lagrangian density \cite{Hat2,Mach}
\begin{eqnarray}
L_{\rho NN} = & &  g_{\rho NN}\left( \bar{\Psi}\gamma_{\mu}\tau^{a}\Psi V_{a}^{\mu} - \frac{\kappa_{\rho}}{2M_{N}}\bar{\Psi}\sigma_{\mu\nu}\tau^{a}\Psi\partial^{\nu}V_{a}^{\mu}\right) ,\nonumber\\
\end{eqnarray}
where $ V_{a}^{\mu} $ is the $ \rho $ meson field and $ \Psi $ is the nucleon 
field. The one-loop contribution to the polarization tensor is
\begin{eqnarray}
\Pi_{\mu\nu}^{\rho NN} = & & 2g_{\rho NN}^{2}\int\frac{d^{4}p}{(2\pi)^{4}}Tr\left[ \Gamma_{\mu}\left( k\right) \frac{1}{\gamma^{\mu}p_{\mu}-M_{N}^{\ast}}\Gamma_{\nu}\left( -k\right) \frac{1}{\gamma^{\nu}\left( p-k\right) _{\nu}-M_{N}^{\ast}}\right] , \nonumber\\
\end{eqnarray}
where
\begin{equation}
\Gamma^{\mu}\left( k\right) = \gamma^{\mu}+\frac{i\kappa_{\rho}}{2M_{N}}\sigma_{\mu\nu}k^{\nu},
\end{equation}
with $ \sigma_{\mu\nu}=\frac{i}{2}[\gamma_{\mu},\gamma_{\nu}] $,
$ M_{N} $ and $ M_{N}^{*} $ are the nucleon masses in vacuum and in the hot hadronic medium
respectively. The polarization tensor $ \Pi_{\mu\nu} ^{\rho NN}\left( k\right) $
can be separated into two parts,
\begin{eqnarray}
\Pi_{\mu\nu} ^{\rho NN}\left( k\right) = & & \left( \frac{k_{\mu}k_{\nu}}{k^{2}}-g_{\mu\nu}\right)\Pi_{F}^{\rho NN}\left( k\right)\
 + \Pi_{D,\mu\nu}^{\rho NN}\left( k\right) ,
\end{eqnarray}
corresponding to the vacuum and the matter contributions.
Using dimensional regularization and taking a phenomenological
subtraction procedure \cite{Hat2}, the vacuum part (T = 0) is,
\begin{eqnarray}
\Pi_{F, L\left( T\right) }^{\rho NN}\left( k\right) = & & k^{2}\left( \frac{g_{\rho NN}}{\pi}\right) ^{2}\left( I_{1}+\frac{\kappa_{\rho}M_{N}^{\ast}}{2M_{N}}I_{2} + \left( \frac{\kappa_{\rho}}{2M_{N}}\right) ^{2}\frac{k^{2}I_{1}+M_{N}^{\ast 2}I_{2}}{2}\right) ,
\end{eqnarray}
where
\begin{eqnarray}
I_{1}=\int_{0}^{1} dx\,x\,\left( 1-x\right) \ln C  ,  I_{2}=\int_{0}^{1} dx \ln C ,   C=\frac{M_{N}^{\ast 2}-x\left( 1-x\right) k^{2}}{M_{N}^{2}-x\left( 1-x\right) k^{2}}.
\end{eqnarray}
The second part $ \Pi_{D,\mu\nu}^{\rho NN}(k) $ is given in terms of the thermal distribution functions
$ n_{N}\left( \mu^{\ast},T\right)  $ and $ \bar{n}_{N}\left( \mu^{*},T\right) $. The longitudinal and the transverse parts of the self-energy can be calculated from $ \Pi_{\mu\nu} $ using equation (15). The longitudinal part of the matter part of the self-energy is given as:
\begin{eqnarray}
\Pi_{D, L}^{\rho NN}\left( k\right) = & &  \Pi_{1D, L}^{\rho NN}\left( k\right) + \Pi_{2D, L}^{\rho NN}\left( k\right) +  \Pi_{3D, L}^{\rho NN}\left( k\right),
\end{eqnarray}
where,
\begin{eqnarray}
\Pi_{1D, L}^{\rho NN}(k) = & &  - \left( \frac{g_{\rho NN}^{2}}{4\pi^{2}}\right) \left( \frac{k^{2}}{|\vec{k}|^{2}}\right)\int_{0}^{\infty} \frac{p^{2}dp}{E_{N}^{*}}\left( n_{N}\left( \mu^{*}, T\right) +\bar{n}_{N}\left( \mu^{*}, T\right) \right)\nonumber\\
& & \left[ 8-\left( \frac{k^{2}+4(E_{N}^{*})^{2}}{p|\vec{k}|}\right) \ln A - \frac{4E_{N}^{*}k_{0}}{p|\vec{k}|}\ln B\right] ,\nonumber\\
\Pi_{2D, L}^{\rho NN}(k) = & & - \left( \frac{g_{\rho NN}^{2}}{\pi^{2}}\right) \left( \frac{k^{2}}{|\vec{k}|}\right)\left( \frac{\kappa_{\rho NN}M_{N}^{*}}{2M_{N}}\right) \int_{0}^{\infty}\frac{pdp}{E_{N}^{*}}\left( n_{N}\left( \mu^{*}, T\right) +\bar{n}_{N}\left( \mu^{*}, T\right) \right) \ln A, \nonumber\\
\Pi_{3D, L}^{\rho NN}(k) = & & \left( \frac{g_{\rho NN}^{2}}{4\pi^{2}}\right) \left( \frac{k^{2}}{|\vec{k}|^{2}}\right)\left( \frac{\kappa_{\rho}^{2}}{4M_{N}^{2}}\right) \int_{0}^{\infty} \frac{p^{2}dp}{E_{N}^{*}}\left( n_{N}\left( \mu^{*}, T\right) +\bar{n}_{N}\left( \mu^{*}, T\right) \right) \nonumber\\
& & \left[  8k_{0}^{2}- \left( \frac{k^{2}k_{0}^{2}+ 4(E_{N}^{*})^{2}k^{2}+4p^{2}|\vec{k}|^{2}}{p|\vec{k}|}\right) \ln A - \frac{4E_{N}^{*}k_{0}^{2}k^{2}}{p|\vec{k}|}\ln B\right] .\nonumber\\
\end{eqnarray}
Here $ p = |\vec{p}| $ and $ E_{N}^{*} $ = $ \sqrt{p^{2}+M_{N}^{*2}} $. $ \vec{k} $
is the 3- momentum of the vector meson, $ \rho $. A and B are defined as
\begin{eqnarray}
A & = & \frac{\left( k^{2}+2p|\vec{k}|\right) ^{2}-4(E_{N}^{*})^{2}k_{0}^{2}}
{\left( k^{2}-2p|\vec{k}|\right) ^{2}-4(E_{N}^{*})^{2}k_{0}^{2}} , \nonumber\\
B & = & \frac{k^{4}-4\left( p|\vec{k}|+E_{N}^{*}k_{0}\right) ^{2}}
{k^{4}-4\left( p|\vec{k}|-E_{N}^{*}k_{0}\right) ^{2}} .
\end{eqnarray} 
Our longitudinal results agree with those of Ref. \cite{Chen}. 
The transverse part of the $ \rho $- meson self energy due to the nucleon loop is given as:
\begin{eqnarray}
\Pi_{D, T}^{\rho NN}\left( k\right) = & &  \Pi_{1D, T}^{\rho NN}\left( k\right) + \Pi_{2D, T}^{\rho NN}\left( k\right) +  \Pi_{3D, T}^{\rho NN}\left( k\right),
\end{eqnarray}
where,
\begin{eqnarray}
\Pi_{1D, T}^{\rho NN}(k) = & &  \left( \frac{g_{\rho NN}^{2}}{4\pi^{2}}\right) \int_{0}^{\infty} \frac{p^{2}dp}{E_{N}^{*}}\left( n_{N}\left( \mu^{*}, T\right) +\bar{n}_{N}\left( \mu^{*}, T\right) \right) \nonumber\\
& & \left[ 4\left( \frac{k_{0}^{2}+ |\vec{k|}^{2} }{|\vec{k}|^{2}}\right) +\left( \frac{|\vec{k}|^{4}-k_{0}^{4}-4(E_{N}^{*})^{2}k^{2}-4p^{2}|\vec{k}|^{2}}{2p|\vec{k}|^{3}}\right) \ln A - \frac{4\k^{2}k_{0}E_{N}^{*}}{2p|\vec{k}|^{3}}\ln B\right] ,\nonumber\\
\Pi_{2D, T}^{\rho NN}(k) = & & - \left( \frac{g_{\rho NN}^{2}}{\pi^{2}}\right) \left( \frac{k^{2}}{|\vec{k}|}\right)\left( \frac{\kappa_{\rho NN}M_{N}^{*}}{2M_{N}}\right) \int_{0}^{\infty}\frac{pdp}{E_{N}^{*}}\left( n_{N}\left( \mu^{*}, T\right) +\bar{n}_{N}\left( \mu^{*}, T\right) \right) \ln A, \nonumber\\
\Pi_{3D, T}^{\rho NN}(k) = & & -k^{2}\left( \frac{g_{\rho NN}^{2}}{4\pi^{2}}\right) \left( \frac{k^{2}}{|\vec{k}|^{2}}\right)\left( \frac{\kappa_{\rho}^{2}}{4M_{N}^{2}}\right) \int_{0}^{\infty} \frac{p^{2}dp}{E_{N}^{*}}\left( n_{N}\left( \mu^{*}, T\right) +\bar{n}_{N}\left( \mu^{*}, T\right) \right) \nonumber\\
& &\left[  4\left( \frac{k^{2}}{|\vec{k}|^{2}}\right) - \left( \frac{k^{4}+ 4k^{2}(E_{N}^{*})^{2}-4p^{2}|\vec{k}|^{2}}{2p|\vec{k}|^{3}}\right) \ln A - \frac{4k^{2}k_{0}E_{N}^{*}}{2p|\vec{k}|^{3}}\ln B\right] .\nonumber\\
\end{eqnarray}

The real and the imaginary parts of the self-energy can be 
obtained after performing the analytic continuation in both
cases as $ k_{0}\rightarrow E + i\varepsilon $ where 
$ E = \sqrt{M_{\rho}^{2}+|\vec{k}|^{2}} $, with $ M_{\rho} $ being
the invariant mass of the $ \rho  $ meson in the medium. The functions $\Pi_{D, L}^{\rho NN}\left( k\right) $ and 
 $\Pi_{D, T}^{\rho NN}\left( k\right) $ acquire an imaginary part when A 
or B are negative.   
This happens when the variable of integration $ p $, lies in the interval,
\begin{eqnarray}
\frac{1}{2}|E\sqrt{1-(4{M_{N}^{*}}^{2})/M_{\rho}^{2}}-|\vec{k}|| 
\leq  p  \leq  \frac{1}{2}|E\sqrt{1-(4{M_{N}^{*}}^{2})/M_{\rho}^{2}}+|\vec{k}||. 
\nonumber\\
\end{eqnarray}
The real and the imaginary parts of the logarithmic function A and B are 
given as:
\begin{eqnarray}
\ln A = \ln |A| - i\pi \Theta(M_{\rho}^{2}-{4M_{N}^{*}}^{2}) \;\;\;\;\;  
{\rm  {and}} \;\;\;\; 
\ln B = \ln |B| + i\pi \Theta (M_{\rho}^{2}-{4M_{N}^{*}}^{2}).\nonumber\\
\end{eqnarray} 

\subsection{$ \rho\pi\pi $ interaction}

With the $ \rho\pi\pi  $ interaction the $ \rho $ meson self-energy is 
calculated by using an effective Lagrangian density describing a system
of charged pions and $ \rho  $ mesons \cite{Gale1}
\begin{eqnarray}
L_{\rho\pi\pi} = & & \frac{1}{2}\mid D_{\mu}\Phi\mid^{2} - \frac{1}{2}m_{\pi}^{2}\mid\Phi\mid^{2} - \frac{1}{4} \rho_{\mu\nu}\rho^{\mu\nu} + \frac{1}{2}m_{\rho}^{2}\rho_{\mu}\rho^{\mu} \nonumber\\
\end{eqnarray}
where $ \Phi $ is the complex charged pion field, 
$ \rho_{\mu\nu} $ = $ \partial_{\mu}\rho_{\nu}-\partial_{\nu}\rho_{\mu} $ is the 
$ \rho  $ field strength, and $ D_{\mu}= \partial_{\mu}- ig_{\rho\pi\pi}\rho_{\mu} $
is the covariant derivative. The one loop contribution to
the polarization tensor is,
\begin{eqnarray}
\Pi_{\mu\nu}\left( k\right) = & & ig_{\rho\pi\pi}^{2}
\int\frac{d^{4}p}{(2\pi)^{4}}\frac{\left( 2p+k\right) _{\mu}\left( 2p+k\right) 
_{\nu}}{\left[ \left( p+k\right) ^{2}-m_{\pi}^{2}\right]\left[ p^{2}-m_{\pi}^{2}\right] } 
- 2i g_{\rho\pi\pi}^{2}\int\frac{d^{4}p}{(2\pi)^{4}} 
\frac{g_{\mu \nu}}{p^{2}-m_{\pi}^{2}} . \nonumber\\
\end{eqnarray}
Separating $ \Pi^{\mu\nu} $ for the $ T = 0 $ and temperature dependent contributions, 
we obtain the contribution for vacuum to be
\begin{eqnarray}
\Pi_{vac}^{\mu\nu}\left( k\right) & = & \left( k^{\mu}k^{\nu}- k^{2}g^{\mu\nu}\right) 
\frac{1}{3}\left( \frac{g_{\rho\pi\pi}}{4\pi}\right) ^{2} \Bigg[ 
\left( 1-\frac{4m_{\pi}^{2}}{k^{2}}\right) ^{\frac{3}{2}} \nonumber\\
& &\ln \left( \frac{\sqrt{1-\frac{4m_{\pi}^{2}}{k^{2}}}
+1}{\sqrt{1-\frac{4m_{\pi}^{2}}{k^{2}}}-1} \right )
-\frac{8m_{\pi}^{2}}{k^{2}} + D\Bigg] 
\end{eqnarray}
where D is the renormalization constant. The renormalization constant D is fixed by $ Re \Pi_{vac}(k^{2}=m_{\rho}^{2}) = 0$ in the free space. This gives D to be 
\begin{eqnarray}
D = -2\left( \frac{m_{\pi}}{\omega_{0}}\right) ^{2}-2\left( \frac{p_{0}}{\omega_{0}}\right) ^{3}\ln \left( \frac{\omega_{0}+p_{0}}{m_{\pi}}\right)  ,
\end{eqnarray} 
where $ 2\omega_{0} = m_{\rho} = 2\sqrt{m_{\pi}^{2}+p_{0}^{2}} $.

The $ T > 0  $ contribution can be split into longitudinal and transverse parts as,
\begin{eqnarray}
\Pi_{L}^{\rho\pi\pi}(k) = - & &\left( \frac{g_{\rho\pi\pi}^{2}}{4\pi^{2}}\right) \left( \frac{k^{2}}{|\vec{k}^{2}|}\right) \int_{0}^{\infty}\frac{p^{2}dp}{\omega}N(\omega)\left[ 4 - \left( \frac{4\omega^{2}+ k_{0}^{2}}{2p|\vec{k}|}\right) \ln a - \left( \frac{2k_{0}\omega}{p|\vec{k}|}\right) \ln b  \right] ,\nonumber\\
\Pi_{T}^{\rho\pi\pi}(k) = & &\left( \frac{g_{\rho\pi\pi}^{2}}{4\pi^{2}}\right)  \int_{0}^{\infty}\frac{p^{2}dp}{\omega}N(\omega)\nonumber\\
& &\left[ \frac{2(k_{0}^2+|{\vec{k}}|^2)}{|{\vec{k}}|^2} - \frac{ (4p^2 +2k_{0}^2-|{\vec{k}}|^2)|{\vec{k}}|^2 - k_{0}^2(k_{0}^2+4\omega^{2})}{4p|{\vec{k}}|^3} \ln a -\left( \frac{k_{0}k^{2}\omega }{p|\vec{k}|^{3}}\right) \ln b \right] \nonumber\\
\end{eqnarray}
where,
\begin{eqnarray}
\omega=\sqrt{p^{2}+m_{\pi}^{2}}  \;\;\;\;  and \;\;\;\;   N(\omega) = \frac{1}{e^{\omega/T}-1} .
\end{eqnarray}
Here $ m_{\pi} $ is the pion mass. In the equation (31), a and b are defined by 
\begin{eqnarray}
 a =  \frac{\left( k^{2}+2p|\vec{k}|\right) ^{2}-4\omega^{2}k_{0}^{2}}
{\left( k^{2}-2p|\vec{k}|\right) ^{2}-4\omega^{2}k_{0}^{2}} , \nonumber\\
b =  \frac{k^{4}-4\left(p|\vec{k}|+\omega k_{0} \right) ^{2}}
{k^{4}-4\left( p|\vec{k}|-\omega {k_{0}}\right) ^{2}} .
\end{eqnarray}

The real and the imaginary parts of the self-energy can be 
obtained as before for the nucleon loop, after performing the analytic continuation as $ k_{0}\rightarrow E + i\varepsilon $ where 
$ E = \sqrt{M_{\rho}^{2}+|\vec{k}|^{2}} $. The $ \rho $ meson
spectral function is obtained as
\begin{eqnarray}
A_{L\left( T\right) }\left( k\right) & = & -2 \frac{Im\Pi_{L\left( T\right) }\left( k\right) }
{\left[ M_{\rho}^{2}-\left( m_{\rho}^{2}+Re\Pi_{L\left( T\right) }\left( k\right) \right)\right] ^{2} + \left[ Im\Pi_{L\left( T\right) }\left( k\right) \right] ^{2} },\nonumber\\  
\end{eqnarray}
with $ \Pi_{L\left( T\right) } $ being the total longitudinal (transverse)
self-energy of $ \rho $ meson.

\section{Results and Discussions}
\label{results}

In this section, we present the results of our calculations of the 
spectral function of $ \rho $-meson in the hot nuclear matter. In our calculations of 
the nuclear properties in the QHD-I model, we have used the 
values for the hadron masses in vacuum and the coupling constants as given in Ref. \cite{Mach}. These values are
$ g_{s}^{2}=109.626 $, $ g_{v}^{2}=190.431 $, $ m_{s}=0.52 $ GeV, $ M_{N}=0.938 $ GeV, $ m_{v}=0.783 $ GeV. 
In the chiral SU(3) model the parameters used are,
$ m_{\pi}=0.1396 $ GeV, $ m_{K}=0.498$ GeV, $ m_{\omega}=0.783$ GeV, 
$f_{\pi}=0.0933$ GeV, $f_{K}=0.122$ GeV, $\zeta_{0}=0.10656$ GeV, $k_{0}= 2.37$, 
$k_{1}= 1.4$, $k_{2}= -5.55$, $k_{3}= -2.64$, $\delta= 2/33 $, $\chi_{0}=0.4027$ 
GeV, $g_{\sigma N}=10.6$, $g_{\zeta N}=-0.47$, $g_{4}=2.7 $ \cite{Zsch2}.
The variation of the effective nucleon mass and effective chemical potential  
determine the spectral function of the rho meson in the hot and dense matter. 
The nuclear matter saturation density is chosen to be
$\rho_{0}=0.16 fm^{-3}$ \cite{Chen} .
The effective nucleon mass and chemical potential are plotted as
functions of temperature for densities $ \rho_{B}  = 0, \rho_{0}, 2\rho_{0}, 4\rho_{0} $ within the frameworks of QHD-I 
and chiral SU(3) in the mean field approximation. The results are shown in figures 1 and 2. Within QHD-I,  for $ \rho_{B}=0 $ as shown in figure 1a, $ M_N^* $ is observed to remain almost a constant till $ T = 0.15 $ GeV above which it is seen to drop to around 0.82 $ M_{N} $ at $ T = 0.2 $ GeV. But within chiral SU(3) model, as shown 
 in figure 2a, $ M_N^* $ is observed to drop to around 0.8 $ M_{N} $ at a temperature, $ T = 0.18 $ GeV. A comparison of figures 1b and 2b shows that the chiral
SU(3) predicts higher values for the nucleon masses and chemical
potential as compared to these in the QHD-I model. In figure 1, with the increase in density $ M_N^* $ is seen to decrease for a given temperature. For a given density, $ M_N^* $ is seen to increase with T till $ T = 0.15 $ GeV above which it decreases with temperature. For a given density, $ \mu^{*} $ is observed to decrease with increase in temperature. Figure 2 follows the same trend. The results show that $ M_{N}^{*} $ and $ \mu^{*} $ in the RHA in both the models are seen to be quite similar to those obtained in the MFA, except that the values with RHA, are seen to be higher than the values of $ M_{N}^{*} $ and $ \mu^{*} $ with  MFA.

\begin{figure}[h]
\includegraphics[width=18cm]{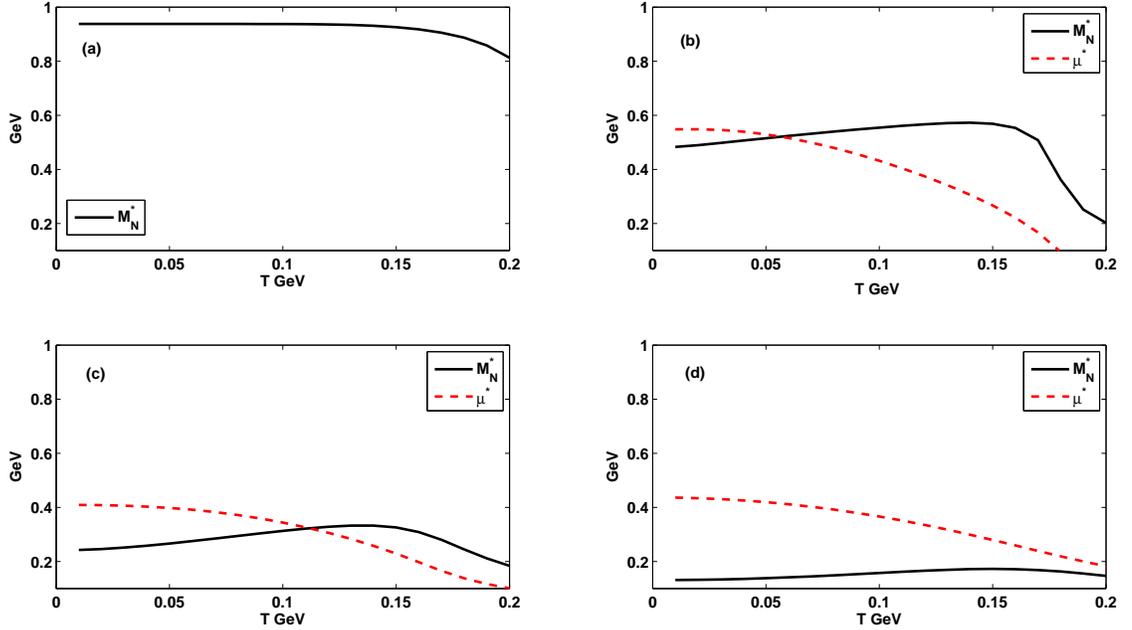} 
\caption{(Color online) Effective nucleon mass $ M_{N}^{\ast} $ and chemical
potential $ \mu^{\ast} $ in MFA for QHD-I model in (a) for $ \rho_{B}=0 $,
(b) for $ \rho_{B}= \rho_{0} $, (c) for $ \rho_{B}= 2\rho_{0} $, (d) for $ \rho_{B}= 4\rho_{0} $ in GeV plotted
as functions of temperature $ T  $ in GeV.} 
\label{fig.1}
\end{figure}

\begin{figure}[h]
\includegraphics[width=18cm]{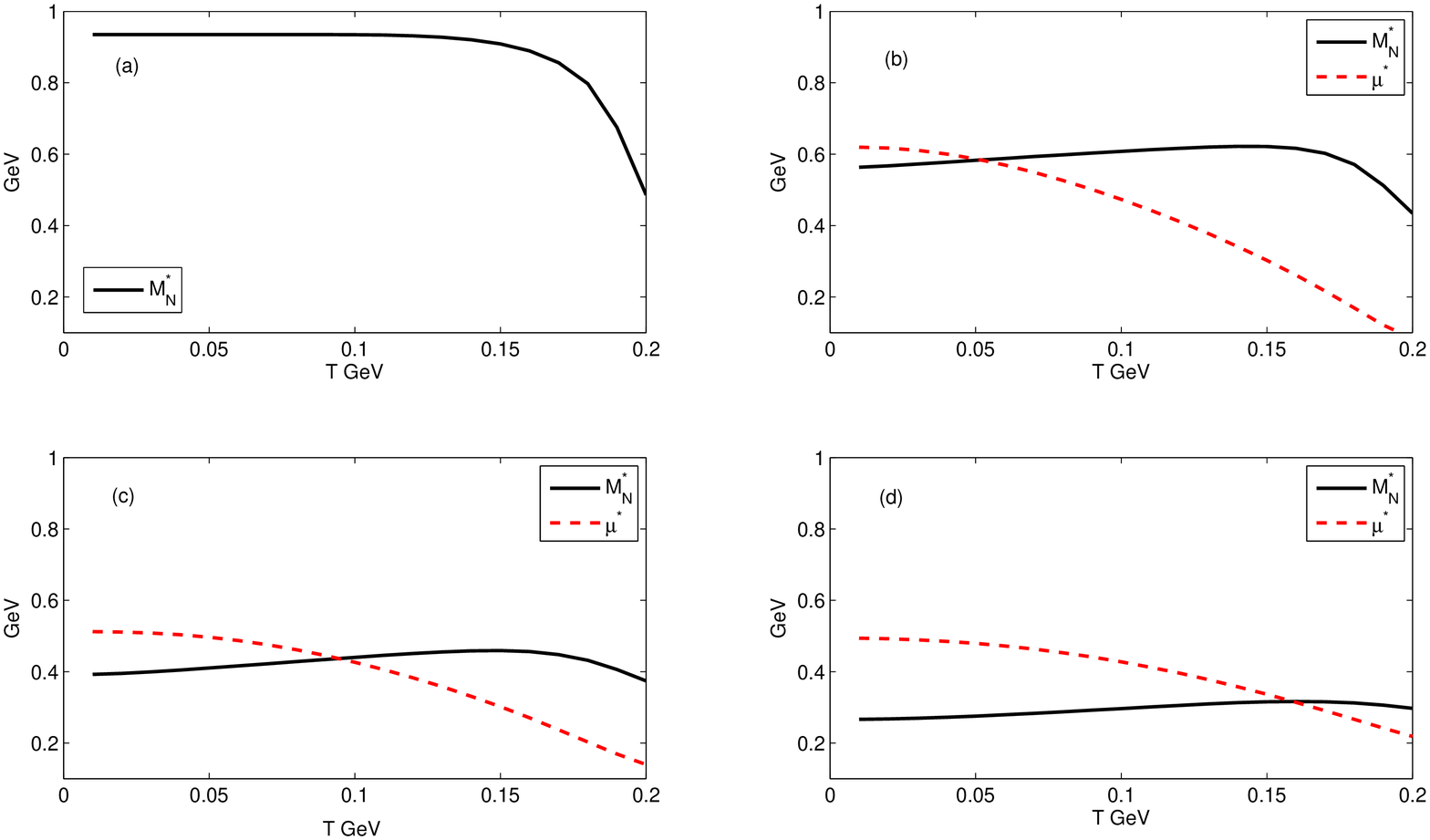} 
\caption{(Color online) Effective nucleon mass $ M_{N}^{\ast} $ and chemical
potential $ \mu^{\ast} $ in MFA for Chiral SU(3) model in (a) for $ \rho_{B}=0 $,
(b) for $ \rho_{B}= \rho_{0} $, (c) for $ \rho_{B}= 2\rho_{0} $, (d) for $ \rho_{B}= 4\rho_{0} $ in GeV plotted
as functions of temperature $ T  $ in GeV.} 
\label{fig.2}
\end{figure}

We consider isospin symmetric case, for which the longitudinal and 
transverse spectral functions are almost the same for all momenta. So for the sake of
convenience, we show here only the longitudinal part of the spectral function. The parameters 
chosen in the calculation of the spectral function are,
$ g_{\rho\pi\pi}^{2}/4\pi= 2.91$, $g_{\rho NN}^{2}=6.96$, $\kappa_{\rho}=6.1 $.
The tensor coupling is very important for the coupling of $ \rho $ meson
with the nucleons. The coupling constants $ g_{\rho NN} $ and $ \kappa_{\rho} $
are determined from the fitting to the nucleon-nucleon scattering
data done by the Bonn group \cite{Mach}.

The longitudinal spectral function is studied for a particular $ \rho $
momentum ( $ |\vec{k}| $ = 0.75 GeV ) for $ \rho_{B} $ = 0 and $ \rho_{B} = \rho_{0} $ at $ T $ = 0.15 GeV in both the QHD-I and chiral SU(3) models.
The results are shown in figure 3. In figure 3a, for $ \rho_{B} $ = 0, the pure temperature effect (only pion loop) is seen to almost coincide with the mean field result in the Walecka  model with the peak around 0.8GeV. With the nucleon loop alone, for small densities $ M_{N}^{*} $ is large and the condition $ M_{\rho}^{2}>4(M_{N}^{*})^{2} $ is not satisfied. Hence there is no imaginary part, which means that there is only a real part to the spectral function implying that  the spectral function is a $ \delta $ function peaked at $ M_{\rho} = 0.77$ GeV. When we consider both the loops within the RHA, it can be seen that the spectral function is seen to shift towards the low mass region and becomes a little sharper due to Dirac sea polarization. The $ \rho $ width decreases because of the decrease in the $ \rho $ mass. Figure 3c for the chiral model, is identical to figure 3a. Here though within MFA the peak is around 0.8GeV, within the RHA, the chiral SU(3) predicts slightly higher value for $ m_{\rho}^{*} $ as a result of higher $ M_{N}^{*} $ values. But for $ \rho_{B} = \rho_{0} $ (figures 3b and 3d), the RHA results for both the models give a spectral function which is further shifted towards the low mass region and is sharper. Here also the chiral model (figure 3d) predicts slightly higher values for $ m_{\rho}^{*} $ and the $ \rho $ width. Within the MFA in both models the peaks are shifted to the high invariant mass region and becomes more wide since $ m_{\rho}^{*} $ increases. This is consistent with the result of Ref. \cite{Zsch1} for $ |\vec{k}| $ = 0. For smaller densities the RHA results are in agreement with the results in Ref. \cite{Chen}.   
\begin{figure}
\includegraphics[width=18cm] {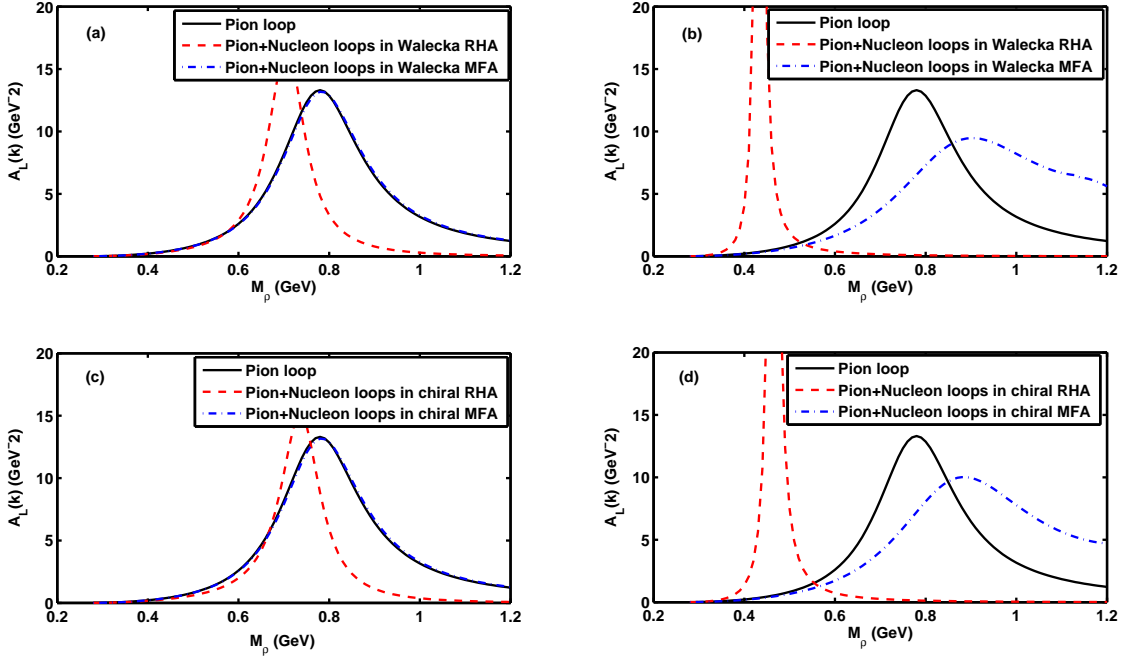} 
\caption{(Color online)Spectral function against the invariant mass $ M_{\rho} $ for $ | \vec{k}| $ = 0.75 GeV and $ T = 0.15 $ GeV for (a) $ \rho_{B} $ = 0 with the pion loop only, the pion and the nucleon loops within the RHA in Walecka model, the pion and the nucleon loops within the MFA in Walecka model, (b) $ \rho_{B} = \rho_{0}  $ with the pion loop only, the pion and the nucleon loops within the RHA in Walecka model, the pion and the nucleon loops within the MFA in Walecka model, (c)  $ \rho_{B} $ = 0 with the pion loop only, the pion and the nucleon loops within the RHA in chiral SU(3) model, the pion and the nucleon loops within the MFA in chiral SU(3) model, and (d) $ \rho_{B} = \rho_{0}  $ with the pion loop only, the pion and the nucleon loops within the RHA in chiral SU(3) model, the pion and the nucleon loops within the MFA in chiral SU(3) model. The coupling constants are $ g_{\rho NN}=6.96$ and $\kappa_{\rho}=6.1 $.} 
\label{fig.3}
\end{figure}

Consider the case when $ \rho_{B} $ = 0.  With the RHA in both the models (figures 4a and 
4c) it can be seen that even though for $ T $ = 0.05GeV and $ T $ = 0.1GeV the peaks coincide at around 
0.77GeV it shifts to around 0.7GeV and 0.72GeV for  $ T $ = 0.15GeV as can be seen in figures 4a and 4b. Hence there is larger reduction in the $ \rho $ mass when the $ M_{N}^{*} $ is calculated with the Walecka model, as compared to in the chiral SU(3) model. Also with RHA the peaks 
become narrower since the $ \rho $ mass decreases. But with MFA as the temperature increases
the peaks take slightly higher $ M_{\rho} $ values in both the models implying a very slight increase in mass and a corresponding increase in the width.  
\begin{figure}
\includegraphics[width=18cm]{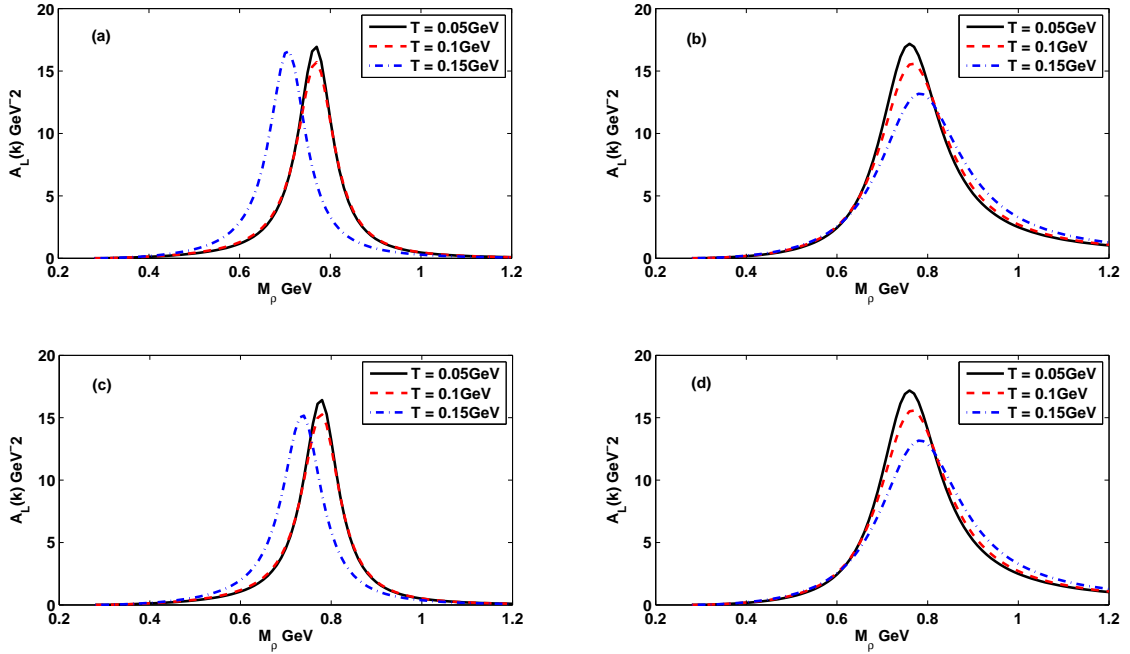} 
\caption{(Color online) Spectral function against the invariant mass $ M_{\rho} $
for $ \rho_{B} = 0 $, $ |\vec{k}| $ = 0.75 GeV for different
values of $ T $ with $ M_{N}^{\ast} $ 
and $ \mu^{\ast} $ calculated with (a)RHA in Walecka,
(b) MFA in Walecka, (c) RHA in chiral SU(3) and (d) MFA in chiral SU(3)models
respectively. The coupling constants are
$ g_{\rho NN}=6.96$ and  $\kappa_{\rho}=6.1 $.} 
\label{fig.4}
\end{figure}

Keeping the 3- momentum of $ \rho $ meson fixed at 0.75GeV, the spectral function
is plotted for various $ T $ values for a higher value of the baryon density $ \rho_{B} = \rho_{0} $ in figure 5. Here again 
figures 5a and 5c are almost identical, figure 5c which corresponds to the RHA in chiral SU(3) model gives slightly higher values for $ m_{\rho}^{*} $ and the $ \rho $ width. In both the models within RHA, $ m_{\rho}^{*} $ and $ \rho $ width increase slightly with temperature. This is consistent with the result in Ref.\cite{Zsch1}. But in figure 5b with MFA and $ M_{N}^{*} $ calculated in the Walecka model the spectral function is observed to have double peaks at $ T $ = 0 and at $ T $ = 0.05 GeV. But at a higher temperature 
the spectral function becomes a smooth curve. The $ \rho $ width is a maximum at $ T $ = 0.1 GeV.
Using MFA, in the chiral SU(3) model, the peak of the spectral function (shown in figure 5d) 
is almost at around the same $ M _{\rho}$ as in Walecka model shown in figure 5b, although in the high invariant mass
region. The $ \rho $ width is observed to be maximum at $ T = 0.05 $ GeV in figure 5d. The spectral function becomes much wider in comparison with the RHA result. 
\begin{figure}
\includegraphics[width=18cm]{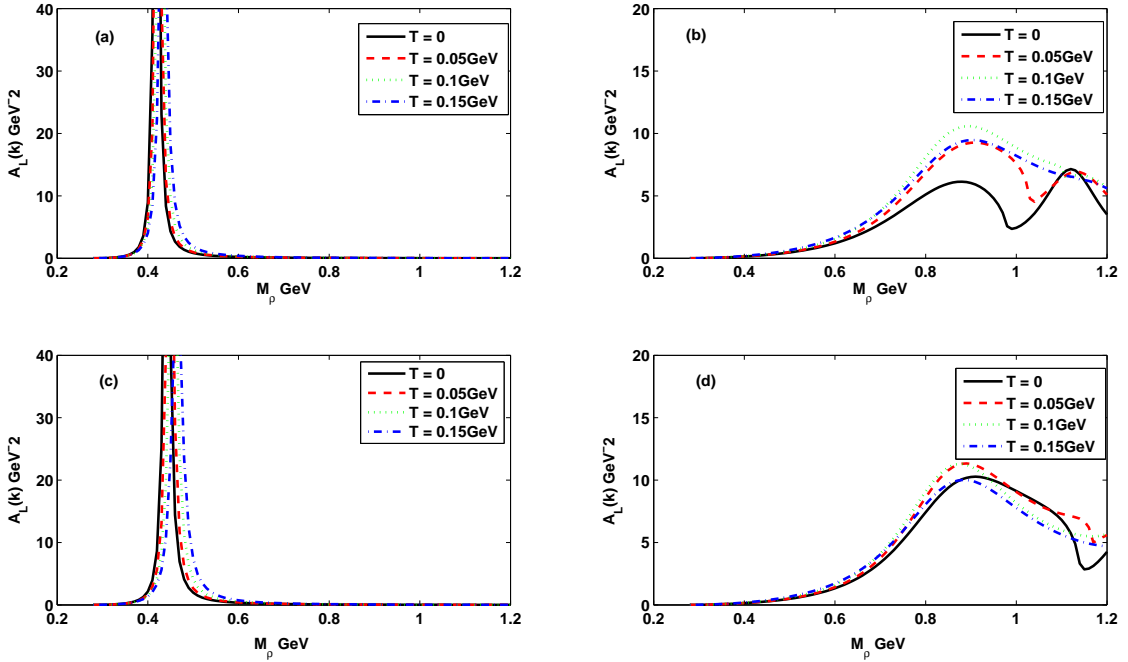} 
\caption{(Color online) Spectral function against the invariant mass $ M_{\rho} $
for $ \rho_{B} = \rho_{0} $, $ | \vec{k}| $ = 0.75 GeV for different
values of $ T $ with $ M_{N}^{\ast} $ 
and $ \mu^{\ast} $ calculated with (a)RHA in Walecka,
(b) MFA in Walecka, (c) RHA in chiral SU(3) and (d) MFA in chiral SU(3)models
respectively. The coupling constants are
$ g_{\rho NN}=6.96$ and $ \kappa_{\rho}=6.1 $.} 
\label{fig.5}
\end{figure}

For $ \rho_{B} = 2\rho_{0} $, the result for the spectral function is shown in figure 6.
In figures 6a and 6c with the RHA calculation in the Walecka and chiral models 
respectively, the plots are observed to be very similar. But a comparison with figures 5a and 5c, which are for a lower density, $ \rho_{B} = \rho_{0} $, 
shows that in figures 6a and 6c the peak  is at a lower $ M_{\rho} $ value for all temperatures. In figure 6a, with the increase in temperature the peaks almost coincide with the peak at $ T = 0 $. In figure 6c there is a very slight increase in $ m_{\rho}^{*} $ with increase in temperature.  
 With MFA in the Walecka model (figure 6b) it can be seen that 
as the density is increased to  $ \rho_{B} = 2\rho_{0} $ the double peaks as seen for $ \rho_{B} = \rho_{0} $ in figure 5b,  merge to a single peak, which is further shifted to the high mass region and the width 
increases with increasing $ T $. This suggests an increase in $ m_{\rho}^{*} $ and a corresponding increase in the width. In figure 6d with MFA in the chiral model, there is a further shift in $ m_{\rho}^{*} $ to the high mass region due to a higher value of $ M_{N}^{*} $.
 \begin{figure}
\includegraphics[width=18cm]{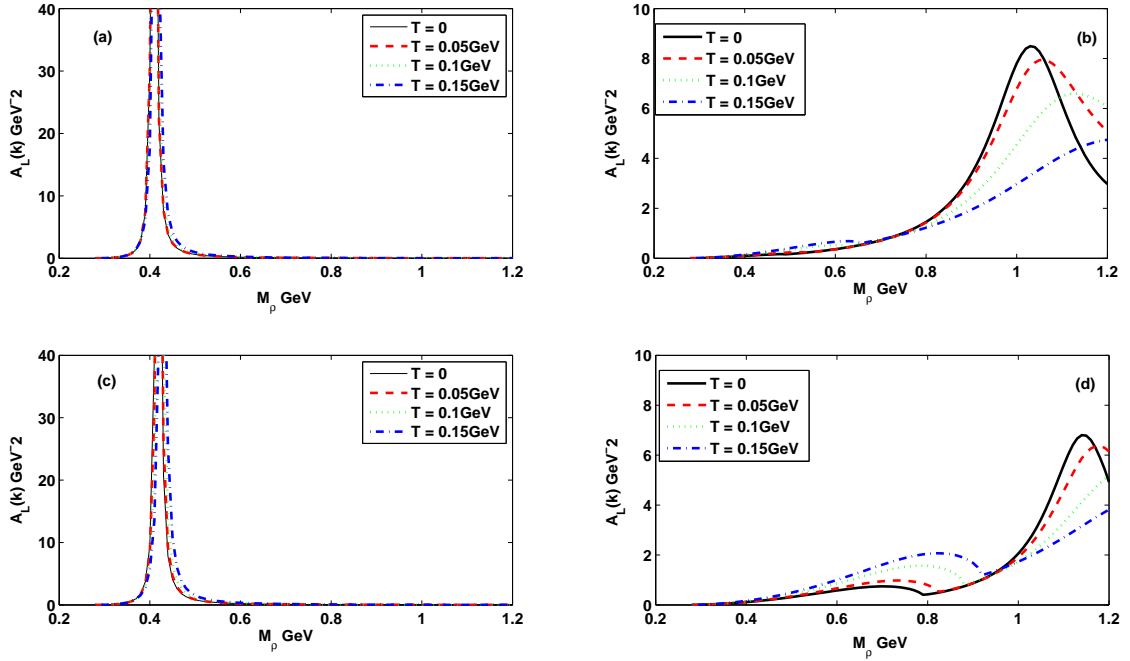} 
\caption{(Color online)Spectral function against the invariant mass $ M_{\rho} $
for $ \rho_{B} = 2\rho_{0} $, $ | \vec{k}| $ = 0.75 GeV for different
values of $ T $ with $ M_{N}^{\ast} $ 
and $ \mu^{\ast} $ calculated with (a)RHA in Walecka,
(b) MFA in Walecka, (c) RHA in chiral SU(3), and (d) MFA in chiral SU(3) models
respectively. The coupling constants are
$ g_{\rho NN}=6.96$ and  $\kappa_{\rho}=6.1 $. } 
\label{fig.6}
\end{figure}

For a higher density of $ \rho_{B} = 4\rho_{0} $, the result for the spectral function is shown in figure 7. A comparison with figure 6 shows that within the RHA in both the models the spectral function peaks at $ T = 0 $ shift slightly to higher $ M_{\rho} $ values  but, with the increase in temperature the peaks almost coincide at a slightly lower value approaching a $ \delta $- like function. A comparison with figures 5 and 6 shows that within RHA, a reduction of the $ \rho $ mass is found upto around $ 2\rho_{0} $. At higher densities, the density dependent part of the $ \rho $ meson self energy, describing the Fermi sea fluctuations starts to be more dominating, leading to slightly increasing masses.Within MFA in the Walecka model (figure 7b), it can be seen that the contribution is small. Since $ M_{N}^{*} $ is very small, at $ \rho_{B} = 4\rho_{0} $, there will be an imaginary part to the $ \rho $ self energy due to the nucleon loop. The inclusion of this reduces the spectral function in our region of interest. But in figure 7d, within MFA in the chiral SU(3) model, $ M_{N}^{*} $ is still too large to give an imaginary part. So again the pion loop alone gives an imaginary part which is responsible for the spectral function in figure 7d.    
\begin{figure}
\includegraphics[width=18cm]{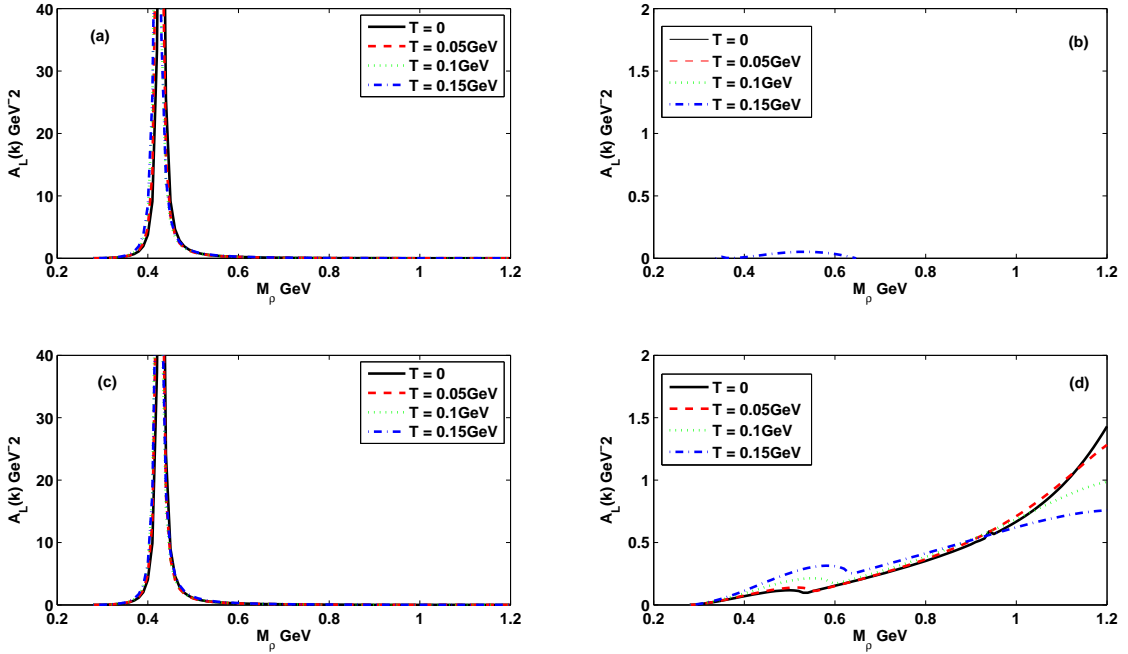} 
\caption{(Color online)Spectral function against the invariant mass $ M_{\rho} $
for $ \rho_{B} = 4\rho_{0} $, $ | \vec{k}| $ = 0.75 GeV for different
values of $ T $ with $ M_{N}^{\ast} $ 
and $ \mu^{\ast} $ calculated with (a)RHA in Walecka,
(b) MFA in Walecka, (c) RHA in chiral SU(3) and (d) MFA in chiral SU(3)models
respectively. The coupling constants are
$ g_{\rho NN}=6.96$ and  $\kappa_{\rho}=6.1 $. } 
\label{fig.7}
\end{figure}

The longitudinal spectral function is studied keeping the density at $ \rho_{B} = 0 $ for different temperatures by changing the $ \rho  $ momentum $ |\vec{k}| $ to 0.25GeV in figure 8.  With the RHA in both the models (figures 8a and 
8c) it can be seen that even though for $ T $ = 0.05GeV and $ T $ = 0.1GeV the peaks coincide at around 
0.77GeV, it shifts to a value around 0.72GeV in figure 8a and around 0.75GeV in figure 8c for  $ T $ = 0.15GeV. Comparison with figures 4a and 4c suggests that when the 3-momentum of the $ \rho $ is decreased the mass drop is smaller. But with MFA, for $ |\vec{k}| $= 0.25GeV, as the temperature increases, the peaks take higher $ M_{\rho} $ values in both the models implying a slight increase in mass and a corresponding increase in the width. Within MFA in both models the finite temperature effects increase with decreasing momentum while with RHA the finite temperature effects decrease with decreasing momentum.    
\begin{figure}
\includegraphics[width=18cm]{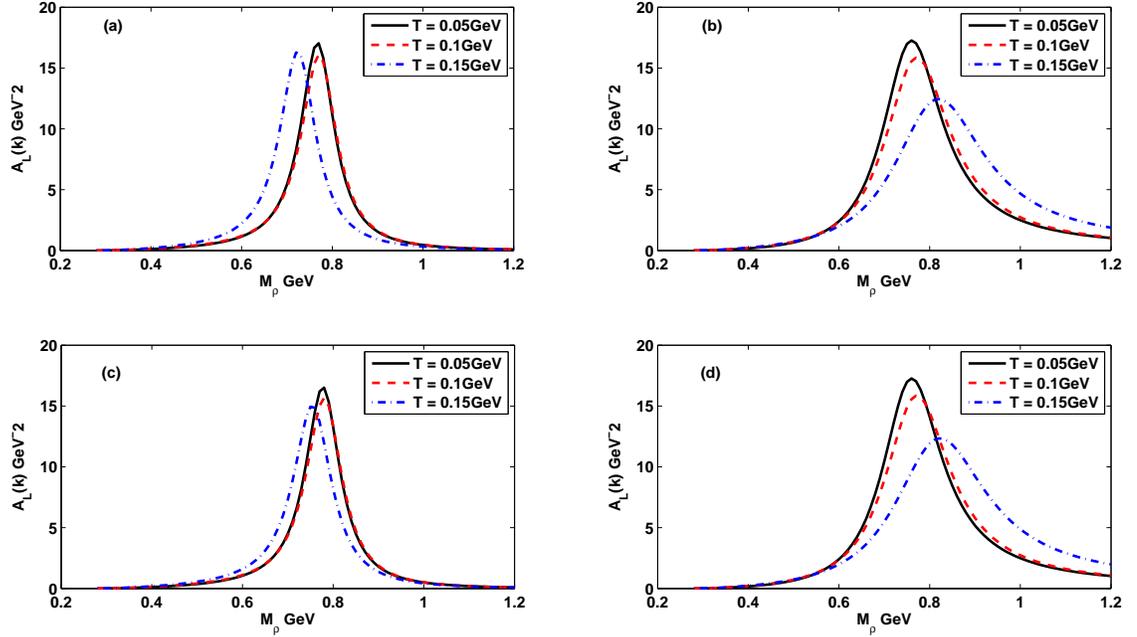} 
\caption{(Color online) Spectral function against the invariant mass $ M_{\rho} $
for $ \rho_{B} = 0 $, $ |\vec{k}| $ = 0.25 GeV for different
values of $ T $ with $ M_{N}^{\ast} $ 
and $ \mu^{\ast} $ calculated with (a)RHA in Walecka,
(b) MFA in Walecka, (c) RHA in chiral SU(3) and (d) MFA in chiral SU(3)models
respectively. The coupling constants are
$ g_{\rho NN}=6.96$ and $\kappa_{\rho}=6.1 $.} 
\label{fig.8}
\end{figure}

For density, $ \rho_{B} = \rho_{0} $, the $ \rho $ momentum fixed at $ |\vec{k}| $ = 0.25GeV, the spectral function is plotted for different temperatures in figure 9. Figure 9a which corresponds to the Walecka model within the RHA gives slightly higher values for the peak position in comparison to the values obtained with $ |\vec{k}| $ = 0.75GeV as shown in figure 5a. The plots for $ T = 0 $, $ T = 0.05 $ GeV, and  $ T = 0.1 $ GeV almost coincide, but the plot for $ T = 0.15 $ GeV shifts to higher $ M_{\rho} $ value and the peak becomes broader suggesting an increase in $ m_{\rho}^{*} $ and the $ \rho $ width. A comparison with figure 8a shows that for the same $ \rho $ 3-momentum and temperature, $ m_{\rho}^{*} $ and $ \rho $ width decrease with increasing value of $ \rho_{B} $. This significant reduction of the $ \rho $ mass due to Dirac sea polarization is found upto around nuclear saturation density. Moreover for the same $ \rho $ 3-momentum, if the density is increased from $ \rho_{B} = 0 $ to $ \rho_{B} = \rho_{0} $, $  m_{\rho}^{*}$ and $ \rho $ width slightly increase with increase in temperature in the latter case, but are observed to decrease with temperature in the former case. Figure 9c corresponding to the chiral SU(3) within the RHA, gives higher values for both $ m_{\rho}^{*} $ and the width than as compared to Walecka model shown in figure 9a. The maximum value of $ m_{\rho}^{*} $      
and width is for $ T = 0.15 $ GeV, the value of $ m_{\rho}^{*} $ being around 0.52GeV. Figure 9b and 9d correspond to the Walecka  and chiral SU(3) models within the MFA. In MFA, $ m_{\rho}^{*} $ and $ \rho $ width tend to increase very much since we are considering only the Fermi sea fluctuations. Our results are similar with the result in Ref. \cite{Zsch1}. Also a comparison with figures 5b and 5d suggests that finite temperature effects are remarkable at low values of 3- momentum of $ \rho $ -mesons in MFA.         
\begin{figure}
\includegraphics[width=18cm]{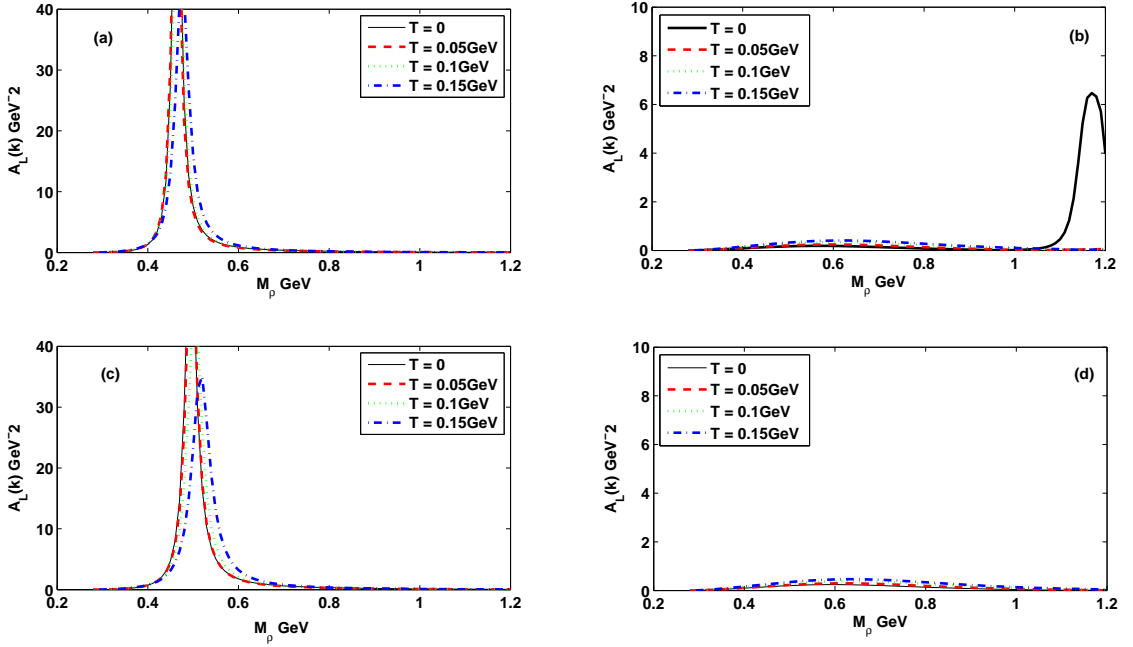} 
\caption{(Color online) Spectral function against the invariant mass $ M_{\rho} $
for $ \rho_{B} = \rho_{0} $, $ | \vec{k}| $ = 0.25 GeV for different
values of $ T $ with $ M_{N}^{\ast} $ 
and $ \mu^{\ast} $ calculated with (a)RHA in Walecka,
(b) MFA in Walecka, (c) RHA in chiral SU(3) and (d) MFA in chiral SU(3)models
respectively. The coupling constants are
$ g_{\rho NN}=6.96$ and  $\kappa_{\rho}=6.1 $.} 
\label{fig.9}
\end{figure}

With the 3- momentum of $ \rho $- meson as 0.25GeV, for $ \rho_{B} = 2\rho_{0} $, the spectral function is plotted for different values of temperature T in figure 10. Comparing figures 10a and 9a corresponding to Walecka model within RHA, it can be seen that at higher density the spectral function follows a different pattern. For $ T = 0 $ the peak is around 0.62GeV. With the increase in temperature the peak shifts further towards low mass region and the width also decreases correspondingly contrary to what is observed in figure 9a. In RHA, a significant reduction of the $ \rho $ mass due to Dirac sea polarization is found up to around nuclear saturation density. But at higher densities, the density dependent part $ \Pi_{D, \mu\nu} $ of the vector meson self energy becomes more dominant, leading to an increase in $ m_{\rho}^{*} $. Comparison of figures 10a and 6a shows that for the same $ \rho_{B} $, $ m_{\rho}^{*} $ and the $ \rho $ width increase with decrease in 3-momentum of the $ \rho $ meson. Figure 10c for the chiral SU(3) model within RHA also follows the same pattern as figure 10a with the peaks taking slightly higher $ M_{\rho} $ values. 
Figures 10b and 10d correspond to the Walecka and the chiral SU(3) models within the MFA. The spectral function contribution is very small in figures 10b and 10d in the region of temperatures we have considered. The peak may be further shifted towards still higher invariant mass region indicating a very high value of $ m_{\rho}^{*} $, outside the range of $ M_{\rho} $ plotted here. Comparison of figures 10b and 6b indicates an increase in the $ \rho $ mass in figure 10b as compared to figure 6b in the Walecka model. The same happens for figures 10d and 6d also for the case of chiral SU(3) model.  
\begin{figure}
\includegraphics[width=18cm]{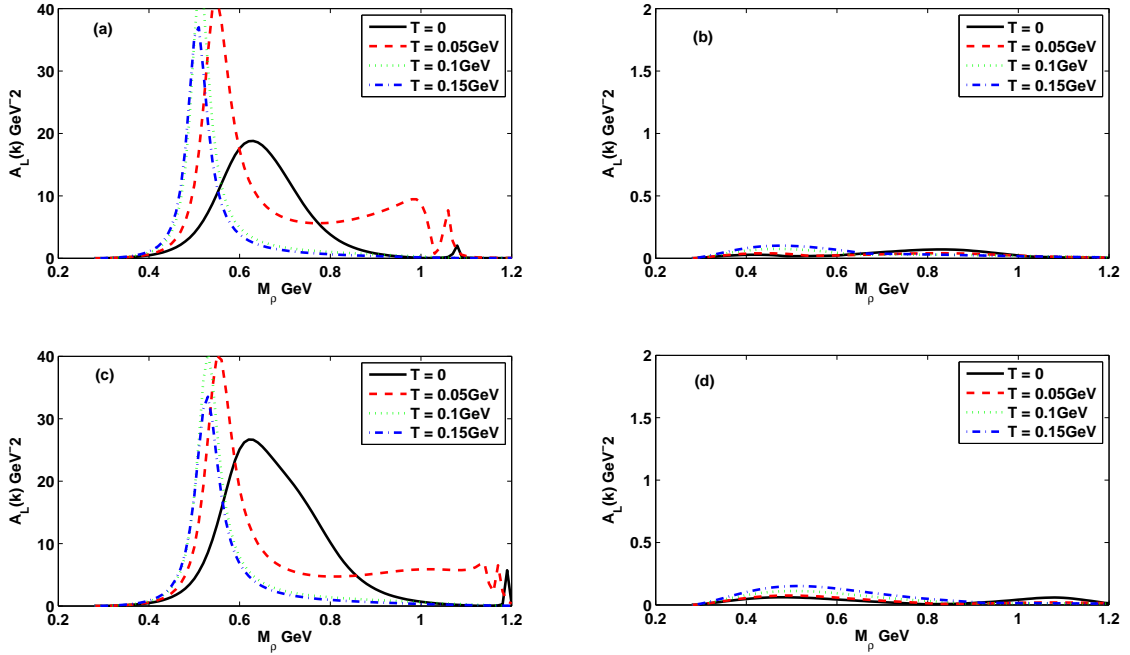} 
\caption{(Color online) Spectral function against the invariant mass $ M_{\rho} $
for $ \rho_{B} = 2\rho_{0} $, $ | \vec{k}| $ = 0.25 GeV for different
values of $ T $ with $ M_{N}^{\ast} $ 
and $ \mu^{\ast} $ calculated with (a)RHA in Walecka,
(b) MFA in Walecka, (c) RHA in chiral SU(3) and (d) MFA in chiral SU(3)models
respectively. The coupling constants are
$ g_{\rho NN}=6.96$ and $\kappa_{\rho}=6.1 $.} 
\label{fig.10}
\end{figure}

With the nucleon density, $ \rho_{B} = 4\rho_{0} $, keeping the $ \rho $ momentum fixed at 0.25GeV, the spectral function is plotted for different temperatures in figure 11. In figures 11a and 11c corresponding to the Walecka and the chiral SU(3) models within RHA, the results are very similar. The spectral function peaks shift to higher $ M_{\rho} $ values and the width increases with increasing temperature. Within MFA in the Walecka and the chiral SU(3) models (figure 11b and 11d) it can be seen that there is no contribution to the spectral function in the region of our interest. Probably the peak may be further shifted towards the high invariant mass region than in figures 10b and 10d.  
\begin{figure}
\includegraphics[width=18cm]{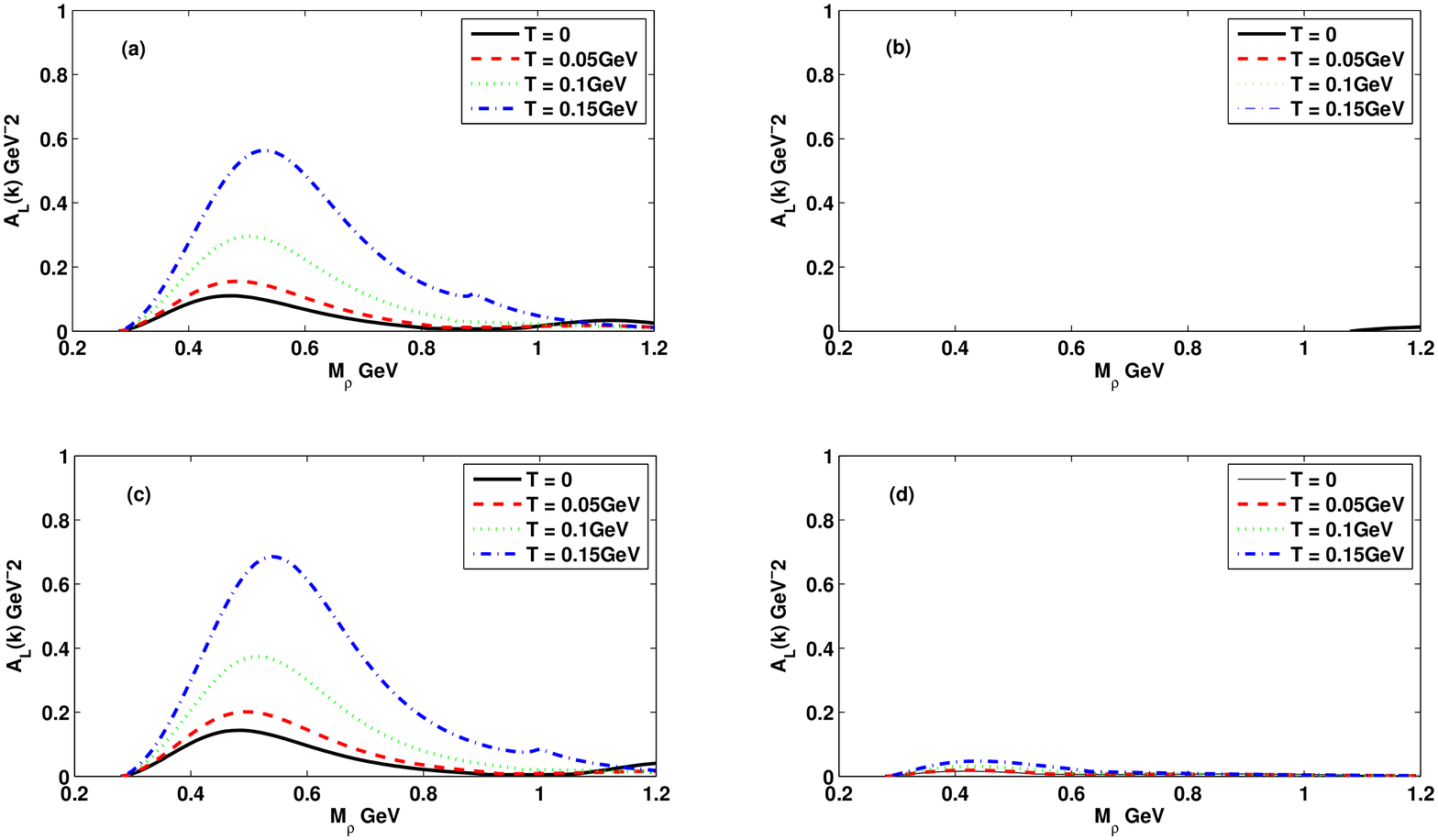} 
\caption{(Color online) Spectral function against the invariant mass $ M_{\rho} $
for $ \rho_{B} = 4\rho_{0} $, $ | \vec{k}| $ = 0.25 GeV for different
values of $ T $ with $ M_{N}^{\ast} $ 
and $ \mu^{\ast} $ calculated with (a)RHA in Walecka,
(b) MFA in Walecka, (c) RHA in chiral SU(3) and (d) MFA in chiral SU(3)models
respectively. The coupling constants are
$ g_{\rho NN}=6.96$ and $\kappa_{\rho}=6.1 $.} 
\label{fig.11}
\end{figure}

So far we have studied the spectral function of the $ \rho $ meson using the $ \rho-N $ coupling strengths as determined
from the $ NN $ forward scattering data. Now we study the $ \rho $ meson spectral function in the chiral
SU(3) model with the nucleon-$ \rho  $ coupling, $ g_{\rho NN} $, as determined
from the symmetry relations \cite{Zsch}. Here we take the tensor coupling as a
parameter in our calculations since this coupling cannot be
fixed from infinite nuclear properties. The coupling constants chosen 
are $ g_{\rho NN}=4.27 $ and $ \kappa_{\rho NN}=2,4,6 $ respectively. Here first we consider the case
when $ |\vec{k}|$  = 0.75GeV and  $ \rho_{B} = \rho_{0} $ and study the spectral function by varying the temperature. The results are shown in figure 12. Figures 12a, 12c and 12e correspond to the the chiral SU(3) model within RHA with the tensor coupling constants as 2,4,6 respectively. In figure 12a where the vector coupling is larger than the tensor coupling we can see that the peak is around 0.5GeV. When the temperature increases the peak slightly shifts indicating a very small increase in $ m_{\rho}^{*} $ and $ \rho $ width.
In figure 12c, the tensor and the vector couplings are comparable. Here it can be seen that the peak is further shifted to the low invariant mass region to a value around 0.4GeV and the behaviour is seen to be similar to figure 12a. Figure 12e correspond to the case where the tensor coupling is dominating over the vector coupling. The peak shifts much towards the low mass region around 0.33GeV and becomes very narrow. Figures 12b, 12dand 12f correspond to the MFA case. In figure 12b, where the vector coupling is dominant almost all the peaks coincide with the peak around 0.82GeV and the $ \rho $ width is also slightly reduced. In figure 12d, the vector and the tensor coupling are comparable. Here we can see that as the temperature is increased the peak is shifted to the high invariant mass region. The $ \rho $ width is maximum at $ T $ = 0.1GeV
above which it is further reduced. In figure 12f, where the tensor coupling is dominant the spectral function gets shifted with temperature again to the high mass region and the width increases with temperature. It can be concluded that within MFA, the $ \rho $ is more stable  when the vector coupling is more  dominant. But within RHA, the $ \rho $ meson is more stable when the tensor coupling is predominant.     
\begin{figure}
\includegraphics[width=18cm]{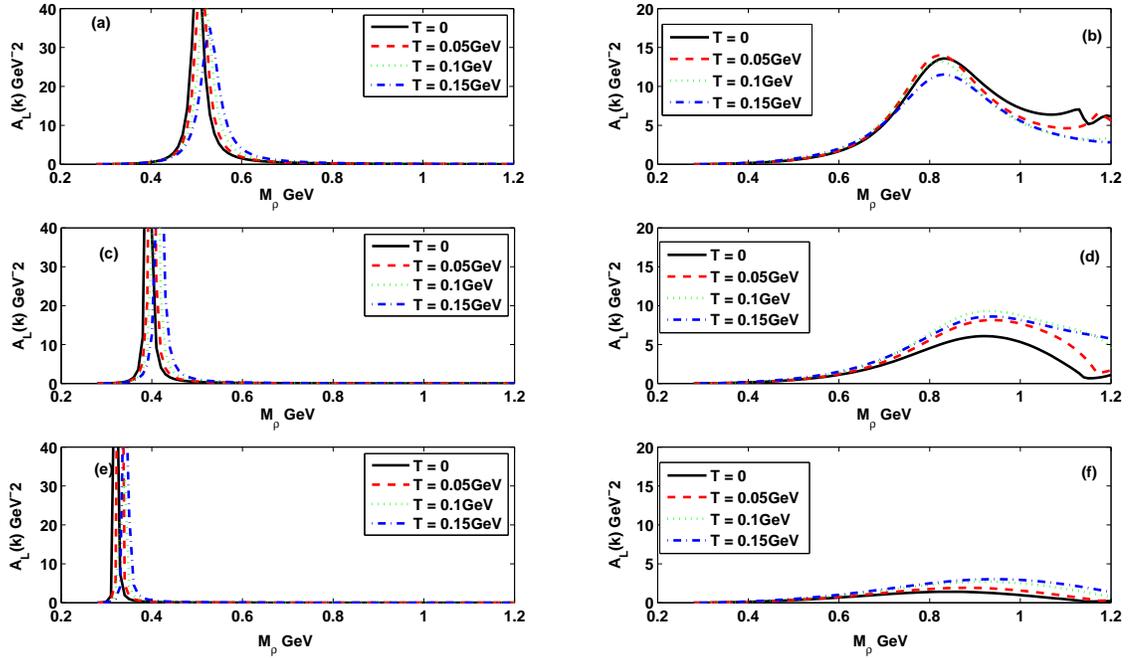} 
\caption{(Color online) Spectral function against the invariant mass $ M_{\rho} $
for $ \rho_{B} = \rho_{0} $, $ |\vec{k}| $ = 0.75 GeV for different
values of $ T $ with (a)RHA with $ \kappa_{\rho NN} $ = 2,
(b) MFA with $ \kappa_{\rho NN} $ = 2, (c) RHA with $ \kappa_{\rho NN} $ = 4, (d) MFA with $ \kappa_{\rho NN} $ = 4, 
(e)RHA with $ \kappa_{\rho NN} $ = 6, (f) MFA with $ \kappa_{\rho NN} $ = 6 
respectively. $ g_{\rho NN} $ = 4.27.} 
\label{fig.12}
\end{figure}

We then study the spectral function varying $ \rho_{B} $ keeping the 
 $ |\vec{k}| $ = 0.75 GeV and $ T $ = 0.15GeV. The results are given in figure 13. Figures 13a, 13c, and 13e correspond to the RHA calculation. With figure 13a ($ \kappa_{\rho NN} $ = 2 ) the peak shifts towards the low mass region as $ \rho_{B} $ increases and the width is also reduced. In figure 13c ($ \kappa_{\rho NN} $ = 4) the reduction in
$ m_{\rho}^{*} $ with  $ \rho_{B} $ is more and the peak is a $ \delta $ function. The $ \rho $ width is also reduced as compared to figure 13a. Figure 13e ($ \kappa_{\rho NN} $ = 6) shows that $ m_{\rho}^{*} $ and $ \rho $ width are reduced very much with increasing $ \rho_{B} $. Figures 13b, 13d and 13f correspond to calculations in MFA. In figure 13b, as  
$ \rho_{B} $ increases the peak is seen to be shifting to the high invariant mass region. In figure 13d ($ \kappa_{\rho NN} $ = 4)  the shift and the width are observed to be larger. For higher densities the contributions are seen to be negligible. With $ \kappa_{\rho NN} $ = 6 the peak is still in the high invariant mass region for  $ \rho_{B} = \rho_{0} $ and the contributions are negligible for higher densities in our region of interest. Within RHA, $ m_{\rho}^{*} $ and the $ \rho $ width decrease with increasing nuclear density whereas within MFA both increase with increasing nuclear density.
\begin{figure}
\includegraphics[width=18cm]{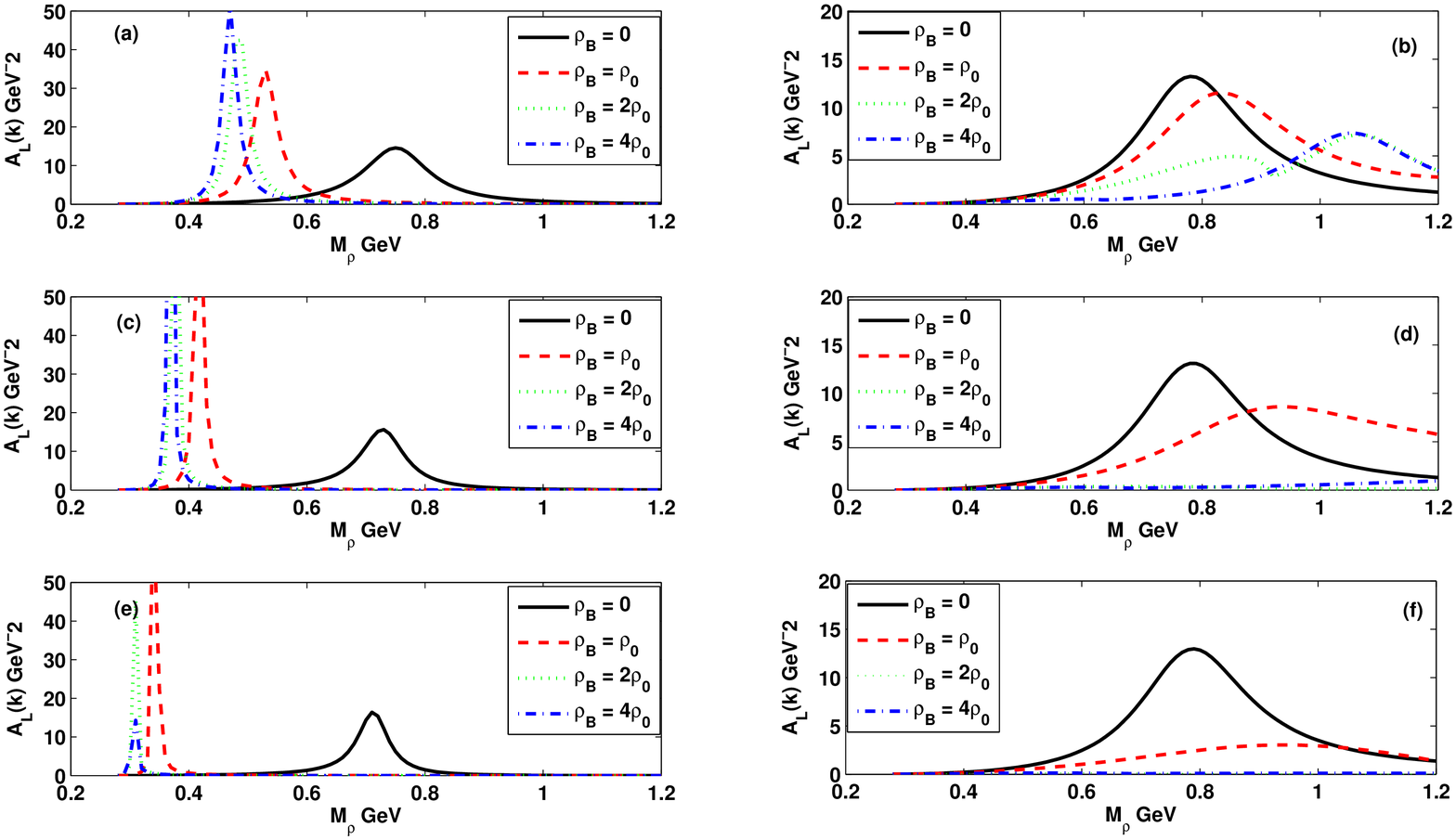} 
\caption{(Color online) Spectral function against the invariant mass $ M_{\rho} $
for  $ T=0.15GeV $, $ |\vec{k}| $  = 0.75 GeV for different
values of the nucleon density $ \rho_{B} $ (a)RHA with $ \kappa_{\rho NN} $ = 2,
(b) MFA with $ \kappa_{\rho NN} $ = 2, (c) RHA with $ \kappa_{\rho NN} $ = 4, (d) MFA with $ \kappa_{\rho NN} $ = 4, 
(e)RHA with $ \kappa_{\rho NN} $ = 6, (f) MFA with $ \kappa_{\rho NN} $ = 6 respectively. $ g_{\rho NN} $ = 4.27.} 
\label{fig.13}
\end{figure}

The spectral function is also studied by varying $ |\vec{k}| $, keeping the temperature and $ \rho_{B} $ fixed. The results are given in figure 14 for $ T=0.15GeV $ and $ \rho_{B} = \rho_{0} $. When $ \kappa_{\rho NN} $ = 2 with RHA (figure 14a) the spectral function has a peak at around 0.62GeV for $ |\vec{k}|$  = 0.25GeV. As the momentum increases the peak shifts and almost remains at the same position at around 0.52GeV. The $ \rho $ width also decreases. With increasing tensor couplings (figure 14c and 14e) the peaks shift further towards the low invariant mass region and become narrower. With MFA, when $ \kappa_{\rho NN} $ = 2, the peak shifts to higher values upto $|\vec{k}|$  = 0.5GeV after which it again falls to around 0.8GeV. When $ \kappa_{\rho NN} $ = 4
$ m_{\rho}^{*} $ increases upto $|\vec{k}|$  = 0.75GeV and at $|\vec{k}|$  = 1GeV the peak is around 0.85GeV and the width is reduced. With $ \kappa_{\rho NN} $ = 6 the shift and the width increase with $|\vec{k}|$ indicating that the particle is highly unstable.           
\begin{figure}
\includegraphics[width=18cm]{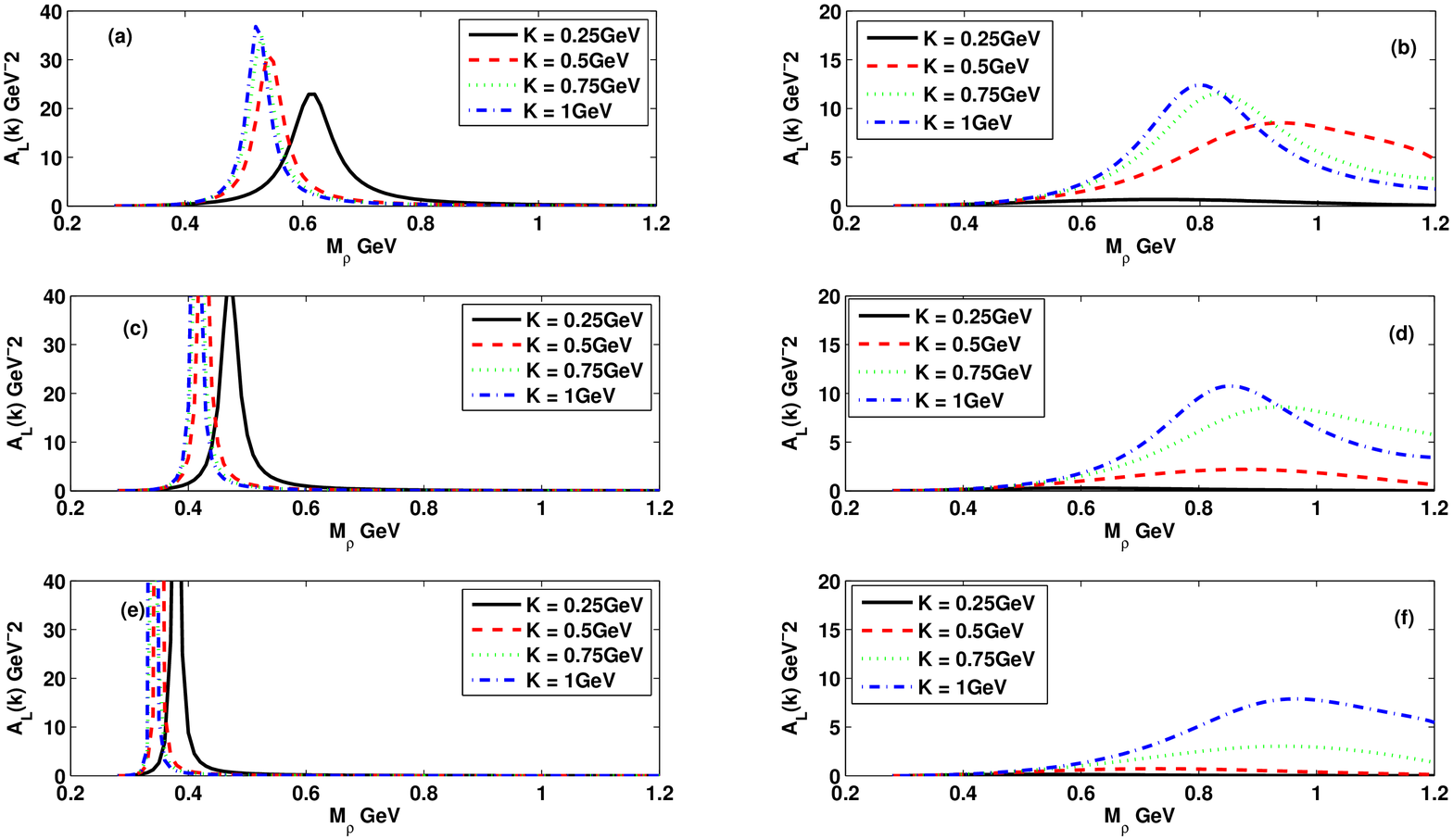} 
\caption{(Color online) Spectral function against the invariant mass $ M_{\rho} $
for  $ T=0.15GeV $, and $ \rho_{B} = \rho_{0} $  for different
values of $ |\vec{k}| $ (a)RHA with $ \kappa_{\rho NN} $ = 2,
(b) MFA with $ \kappa_{\rho NN} $ = 2, (c) RHA with $ \kappa_{\rho NN} $ = 4, (d) MFA with $ \kappa_{\rho NN} $ = 4, 
(e)RHA with $ \kappa_{\rho NN} $ = 6, (f) MFA with $ \kappa_{\rho NN} $ = 6 respectively. $ g_{\rho NN} $ = 4.27.} 
\label{fig.14}
\end{figure}

\section{summary}

To summarize, in the present work, we have investigated  the temperature and density effects
on the $ \rho $ meson spectral function with the effective lagrangian
in the ambit of the QHD-I model and the chiral SU(3) model in isospin symmetric pion and nucleon media. With only the pure temperature
effect, i.e., with pion loop only, the medium corrections are observed to be modest 
even upto a temperature, T = 0.15 GeV. But the inclusion of the nucleon loop
drastically changes the medium properties of the $ \rho $ meson, meaning
that the density effect is a remarkable effect as compared to the pure
temperature effect. In QHD-I model, $ M_{N}^{\ast} $ has a larger drop with density as compared to the chiral SU(3) model. This is 
reflected in the spectral function of the $ \rho $ meson. 
Even for small densities the spectral function is shifted towards 
the low invariant mass region significantly and the spectral function
becomes very sharp in the RHA. For the same density the spectral function gets shifted slightly towards the high invariant mass region and becomes broad in the MFA. This is consistent with the recent experimental result which supports a broadening induced in the medium.   

Also it can be seen that the spectral function depends on the 
coupling strengths of the $ \rho $ meson with the nucleons.
If the tensor coupling is dominant then $ m_{\rho}^{\ast} $ has
a tendency to take comparatively lower values. If the vector coupling
is predominant then $ m_{\rho}^{\ast} $ has a tendency to take slightly
higher values in RHA. The $ \rho $ width also follows the same pattern.
But in the MFA when the vector coupling is dominant $ m_{\rho}^{\ast} $ and $ \rho $ width take comparatively lower values than
when the tensor coupling is dominant.

The spectral function of the $ \rho $ meson also depends on its
momentum {$ |\vec{k}| $}. For the same temperature and baryon density, $ \rho_{B} $, the spectral 
function shifts to the low mass region indicating a drop in
$ m_{\rho}^{\ast} $ with {$ |\vec{k}| $} in RHA irrespective of whether the vector coupling or the tensor coupling is dominant. In MFA for lower momenta the peak is at a higher $ M_{\rho} $
value, but as the momentum increases the peak takes a comparatively lower value though in the high invariant mass region . 

\acknowledgements
The authors gratefully acknowledge financial support from Department of Science \& Technology, Government of India, for the projects SR/WOS-A/PS/10/2007 and SR/S2/HEP-21/2006. AM would like to acknowledge Alexander von Humboldt foundation for financial support and FIAS, University of Frankfurt for warm hospitality when this work was initiated.

\end{document}